\documentclass[12pt,a4paper]{article}
\usepackage{a4wide,graphics,amssymb}
\usepackage[intlimits]{amsmath}
\numberwithin{equation}{section}

\newcommand{\Q}{\widetilde{Q}}
\newcommand{\D}{\rlap{\hspace{0.2em}/}D}
\newcommand{\p}{\widetilde{p}}
\renewcommand{\S}{\widetilde{S}}
\newcommand{\Z}{\widetilde{Z}}
\newcommand{\Si}{\widetilde{\Sigma}}
\newcommand{\La}{\widetilde{\Lambda}}
\newcommand{\Ga}{\widetilde{\Gamma}}
\newcommand{\ga}{\widetilde{\gamma}}

\newcommand{\MS}{\ensuremath{\overline{\mathrm{MS}}}}
\newcommand{\Tr}{\mathop{\mathrm{Tr}}\nolimits}
\newcommand{\re}{\mathop{\mathrm{Re}}\nolimits}

\newcommand{\1}[1]{\mathbf{1}^#1}
\newcommand{\2}[1]{\mathbf{2}^#1}
\newcommand{\3}[1]{\mathbf{3}^#1}
\newcommand{\4}[1]{\mathbf{4}^#1}
\newcommand{\5}[1]{\mathbf{5}^#1}

\begin{document}

\begin{flushright}
TTP00-19
\end{flushright}
\vspace{1cm}
\begin{center}
\Huge Lectures on perturbative HQET 1\\[5mm]
\normalsize A.~G.~Grozin\\
Institut f\"ur Theoretische Teilchenphysik, Universit\"at Karlsruhe
\end{center}

\begin{abstract}
Extended version of lectures given at the University of Karlsruhe
and at Calc-2000 school in Dubna.
Properties of HQET as a field theory, methods of calculation of HQET diagrams
and some simple applications are explained in detail.
These lectures can be used as an additional chapter with any modern QCD textbook.
\end{abstract}

\section{HQET Lagrangian}
\label{SecHQET}

Let's consider QCD with a heavy flavour $Q$ with mass $m$
and a number of light flavours.
We shall be interested in problems with a single heavy quark
staying approximately at rest.
More exactly, let $\omega\ll m$ be the characteristic momentum scale.
We shall assume that the heavy quark has the momentum $|\vec{p}\,|\lesssim\omega$
and the energy $|p_0-m|\lesssim\omega$;
light quarks and gluons have momenta $|\vec{k}_i|\lesssim\omega$
and energies $|k_{0i}|\lesssim\omega$.
Heavy quark effective theory (HQET) is an effective field theory
constructed to reproduce QCD results for such problems
expanded up to some order in $\omega/m$.
In practice, only a few first orders in $1/m$ expansion are considered,
because complexity of the theory grows fast with the order.

Let's start from the QCD Lagrangian
\begin{equation}
L = \overline{Q}(i\D-m)Q + \cdots
\label{QCDlagr}
\end{equation}
where $Q$ is the heavy quark field,
and dots mean all the terms with light quarks and gluons.
The free heavy quark Lagrangian $\overline{Q}(i\rlap/\partial-m)Q$
gives the dependence of the energy
on the momentum $p_0=\sqrt{m^2+\vec p\,^2}$.
If we assume that characteristic momenta $|\vec p\,|\ll m$,
then we can simplify the dispersion law to $p_0=m$.
It corresponds to the Lagrangian $\overline{Q}(i\gamma_0\partial_0-m)Q$.
In our class of problems, the lowest-energy state (``vacuum'')
consists of a single particle --- the heavy quark at rest.
Therefore, it is convenient to use the energy of this state $m$
as a new zero level.
This means that instead of the true energy $p_0$
of the heavy quark (or any state containing this quark)
we shall use the residual energy $\p_0=p_0-m$.
Then the on-shell heavy quark has the energy $\p_0=0$
independently on the momentum.
The free Lagrangian giving such a dispersion law
is $\overline{Q}i\gamma_0\partial_0 Q$.
The spin of the heavy quark at rest can be described
by a 2-component spinor $\Q$
(we can also consider it as a 4-component spinor
with the vanishing lower components: $\gamma_0\Q=\Q$).
Reintroducing the interaction with the gluon field by requirement of the gauge invariance,
we arrive at the HQET Lagrangian~\cite{EH1}
\begin{equation}
L = \Q^+ iD_0 \Q + \cdots
\label{HQETlagr}
\end{equation}
where all light-field parts (denoted by dots)
are exactly the same as in QCD.
The field theory~(\ref{HQETlagr}) is not Lorentz-invariant,
because the heavy quark defines a selected frame --- its rest frame.

The Lagrangian~(\ref{HQETlagr}) gives the static quark propagator
\begin{equation}
\S(\p) = \frac{1}{\p_0+i0}\,,\quad
\S(x) = \S(x_0)\delta(\vec x)\,,\quad
\S(t) = -i\theta(t)\,.
\label{HQETprop}
\end{equation}
In the momentum space it depends only on $\p_0$ but not on $\vec p$,
because we have neglected the kinetic energy.
Therefore, in the coordinate space the static quark does not move.
The unit $2\times2$ matrix is assumed in the propagator~(\ref{HQETprop}).
It is often convenient to use it as a $4\times4$ matrix;
in such a case the projector $\frac{1+\gamma_0}{2}$
onto the upper components is assumed.
The static quark interacts only with $A_0$;
the vertex is $ig\delta_0^\mu t^a$.

The static quark propagator in a gluon field
is given by the straight Wilson line
\begin{equation}
\S(x) = -i\theta(x_0)\delta(\vec{x}) P\exp ig\int A_\mu dx^\mu\,.
\label{WilsonLine}
\end{equation}
Many properties of HQET were first derived in the course of investigation
of renormalization of Wilson lines in QCD.
In fact, the HQET Lagrangian was used as a technical device
for investigation of Wilson lines.

Loops of a static quark vanish, because it propagates only forward in time.
In other words, in the momentum space, all poles in the $\p_0$ plane
lie in the lower half-plane; closing integration contours upwards, we get zero.

The Lagrangian~(\ref{HQETlagr}) can be rewritten in covariant notations~\cite{Gr,Ge}:
\begin{equation}
L_v = \overline{\Q}_v iv\cdot D \Q_v + \cdots
\label{CovarLagr}
\end{equation}
where the static quark field $\Q_v$ is a 4-component spinor
obeying the relation $\rlap/v\Q_v=\Q_v$,
and $v^\mu$ is the quark velocity.
The momentum $p$ of the heavy quark (or any state containing it)
is related to the residual momentum $\p$ by
\begin{equation}
p=mv+\p\,, \quad |\p^\mu|\ll m\,.
\label{HQETp}
\end{equation}
The static quark propagator is
\begin{equation}
\S(\p) = \frac{1+\rlap/v}{2} \frac{1}{\p\cdot v+i0}\,,
\label{CovarProp}
\end{equation}
and the vertex is $igv^\mu t^a$.

One can watch~\cite{Gr} how expressions for QCD diagrams tend to
the corresponding HQET expressions in the limit $m\to\infty$.
The QCD heavy quark propagator is
\begin{equation}
S(p) = \frac{\rlap/p+m}{p^2-m^2}
     = \frac{m(1+\rlap/v)+\rlap/\p}{2m\p\cdot v+\p\,^2}
     = \frac{1+\rlap/v}{2}\frac{1}{\p\cdot v} + \mathcal{O}\left(\frac{\p}{m}\right)\,.
\label{QCDprop}
\end{equation}
A vertex $ig\gamma^\mu t^a$ sandwiched between two projectors
$\frac{1+\rlap/v}{2}$ may be replaced by $ig v^\mu t^a$
(one may insert the projectors at external heavy quark legs, too).
Therefore, any tree QCD diagram equals the
corresponding HQET one up to $\mathcal{O}(\p/m)$ terms.
In loops, momenta can be arbitrarily large,
and the relation~(\ref{QCDprop}) can break.
But if we renormalize HQET, loop integrals become convergent.
In them, characteristic loop momenta are of order $\omega$,
and one may use~(\ref{QCDprop}).

Renormalization properties (anomalous dimensions etc.) of HQET
differ from those of QCD.
The ultraviolet behavior of an HQET diagram is determined
by the region of loop momenta much larger than the characteristic momentum scale
of the process $\omega$, but much less than the heavy quark mass $m$
(which tends to infinity from the very beginning).
It has nothing to do with the ultraviolet behavior
of the corresponding QCD diagram with the heavy quark line,
which is determined by the region of loop momenta much larger than $m$.
In the conventional QCD, the first region produces hybrid logarithms
which are difficult to sum up.
In HQET, hybrid logarithms become ultraviolet logarithmic divergences
governed by the renormalization group with corresponding anomalous dimensions.

The HQET Lagrangian~(\ref{HQETlagr})
possesses the $SU(2)$ heavy quark spin symmetry~\cite{IW}.
The heavy quark spin does not interact with gluon field at $m\to\infty$,
because its chromomagnetic moment is of the order of $1/m$ by dimensionality.
Therefore, we can rotate the heavy quark spin separately.

Not only the orientation, but also the magnitude
of the heavy quark spin is irrelevant in HQET.
We can switch off the heavy quark spin, making it spinless.
This often simplifies counting form factors in HQET.
Or, if we want, we can make the heavy quark to have spin 1.
This leads to a supersymmetry group called the superflavour symmetry~\cite{Super}.
It allows one to predict properties of hadrons with a scalar or vector
heavy quark (appearing in some extensions of the Standard Model).
This idea can also be applied to baryons with two heavy quarks,
because they form a small size (of order $1/(m\alpha_s)$) bound state
with spin 0 or 1, antitriplet in colour.

HQET has great advantages over QCD
in lattice simulation of heavy quark problems.
Indeed, the applicability conditions of the lattice approximation
to problems with light hadrons require
that the lattice spacing is much less than the characteristic hadron size,
and the total lattice length is much larger than this size.
For simulation of QCD with a heavy quark,
the lattice spacing must be much less than the heavy quark Compton wavelength $1/m$.
For $b$ quark, this is technically impossible at present.
The HQET Lagrangian does not involve the heavy quark mass $m$,
and the applicability conditions of the lattice approximation to HQET
are the same in the case of light hadrons.

\section{One-loop massless propagator diagrams}
\label{SecQCD1}

The main aim of these lectures is to discuss the methods used for calculation
of Feynman diagrams in HQET, and some of the simplest applications.
These methods have much in common with the ones used to calculate
multiloop diagrams in massless theories.
Therefore, for convenience of the readers,
we shall recall some well-known facts about massless diagrams, too.

\begin{figure}[ht]
\begin{center}
\begin{picture}(32,20)
\put(16,10){\makebox(0,0){\includegraphics{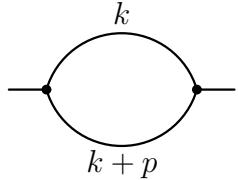}}}
\put(16,0){\makebox(0,0){$k+p$}}
\put(16,20){\makebox(0,0){$k$}}
\end{picture}
\end{center}
\caption{One-loop massless propagator diagram}
\label{L1}
\end{figure}

In this Section, we shall calculate the one-loop massless propagator diagram
with arbitrary degrees of the denominators (Fig.~\ref{L1})
\begin{equation}
G = \int \frac{d^d k}{(-(k+p)^2-i0)^{n_1}(-k^2-i0)^{n_2}}\,,
\label{l1}
\end{equation}
where $d=4-2\epsilon$ is the space-time dimension.

The first step is to combine the denominators together.
To this end, we write $1/a^n$ for $\re a>0$ as
\begin{equation}
\frac{1}{a^n} = \frac{1}{\Gamma(n)}
\int_0^\infty e^{-a\alpha}\, \alpha^{n-1}\, d\alpha
\label{alpha}
\end{equation}
($\alpha$-representation).
Multiplying two such representations, we have
\begin{equation}
\frac{1}{a_1^{n_1} a_2^{n_2}} = \frac{1}{\Gamma(n_1) \Gamma(n_2)}
\int e^{- a_1 \alpha_1 - a_2 \alpha_2}\, \alpha_1^{n_1-1} \alpha_2^{n_2-1}\,
d\alpha_1\, d\alpha_2\,.
\label{alpha2}
\end{equation}
Now we proceed to the new variables $\alpha_1=x\alpha$, $\alpha_2=(1-x)\alpha$,
and obtain
\begin{equation}
\frac{1}{a_1^{n_1} a_2^{n_2}} = \frac{\Gamma(n_1+n_2)}{\Gamma(n_1) \Gamma(n_2)}
\int_0^1 \frac{x^{n_1-1} (1-x)^{n_2-1}\, dx}
{\left[a_1 x + a_2 (1-x)\right]^{n_1+n_2}}\,.
\label{Feyn}
\end{equation}
This Feynman parametrization is valid not only when $\re a_{1,2}>0$,
but, by the analytical continuation,
in all cases when the integral in $x$ is well-defined.

Combining the denominators in~(\ref{l1})
and shifting the integration momentum $k\to k-xp$, we obtain
\begin{equation}
G = \frac{\Gamma(n_1+n_2)}{\Gamma(n_1) \Gamma(n_2)}
\int \frac{x^{n_1-1} (1-x)^{n_2-1}\, dx\, d^d k}
{\left[-k^2+x(1-x)(-p^2)-i0\right]^{n_1+n_2}}\,.
\label{l1Comb}
\end{equation}

Now we shall calculate the one-loop massive vacuum diagram,
which appear as a sub-expression in many calculations.
Making the Wick rotation $k_0=ik_{E0}$ and going to the Euclidean space
$k^2=-k_E^2$, we have
(here $2\pi^{d/2}/\Gamma(d/2)$ is $d$-dimensional full solid angle)
\begin{equation}
\int \frac{d^d k}{(-k^2+m^2-i0)^n} = i \frac{2\pi^{d/2}}{\Gamma(d/2)}
\int \frac{k_E^{d-1}\, d k_E}{(k_E^2+m^2)^n}\,.
\label{Tadpole0}
\end{equation}
At this point, the meaning of the $d$-dimensional integral
for an arbitrary $d$ becomes completely well-defined.
Then we use $2 k_E\, d k_E \to d k_E^2$,
and proceed to the dimensionless variable $z=k_E^2/m^2$ to obtain
\begin{equation*}
\frac{i\pi^{d/2}}{\Gamma(d/2)} (m^2)^{d/2-n}
\int_0^\infty \frac{z^{d/2-1}\, dz}{(z+1)^n}\,.
\end{equation*}
The substitution $x=1/(z+1)$ reduces this integral to the Euler B-function,
and we arrive at
\begin{equation}
\int \frac{d^d k}{(-k^2+m^2-i0)^n} =
i \pi^{d/2} \frac{\Gamma(-d/2+n)}{\Gamma(n)} (m^2)^{d/2-n}\,.
\label{Tadpole}
\end{equation}

We can now resume the calculation of the integral $G$~(\ref{l1Comb}).
We shall assume that $p^2<0$, so that production of a pair
of on-shell massless particles is not possible.
Using~(\ref{Tadpole}) with $m^2\to x(1-x)(-p^2)$ and $n\to n_1+n_2$, we obtain
\begin{equation}
G = i \pi^{d/2} \frac{\Gamma(-d/2+n_1+n_2)}{\Gamma(n_1) \Gamma(n_2)}
(-p^2)^{d/2-n_1-n_2} \int_0^1 x^{d/2-n_2-1} (1-x)^{d/2-n_1-1}\, dx\,.
\label{l1Tad}
\end{equation}
This integral is a B-function.
Our final result can be written as
\begin{equation}
\begin{split}
& \int \frac{d^d k}{D_1^{n_1} D_2^{n_2}} =
i \pi^{d/2} (-p^2)^{d/2-n_1-n_2} G(n_1,n_2)\,,\\
& D_1 = -(k+p)^2\,, \quad D_2 = -k^2\,,\\
& G(n_1,n_2) = \frac{\Gamma(-d/2+n_1+n_2) \Gamma(d/2-n_1) \Gamma(d/2-n_2)}
{\Gamma(n_1) \Gamma(n_2) \Gamma(d-n_1-n_2)}\,,
\end{split}
\label{G1}
\end{equation}
where the correct $i0$ terms are assumed in $D_{1,2}$.
If $n_{1,2}$ are integer, $G(n_1,n_2)$ is proportional to
$G_1=\Gamma(1+\epsilon)\Gamma^2(1-\epsilon)/\Gamma(1-2\epsilon)$,
the coefficient being a rational function of $d$.

Ultraviolet (UV) divergences of the original integral~(\ref{l1})
are reproduced by the loop-momentum integral~(\ref{Tadpole}),
and are given by the poles of $\Gamma(-d/2+n_1+n_2)$
(with minus sign in front of $d$).
Infrared (IR) divergences of the original integral reside in the integral
in the Feynman parameter $x$~(\ref{l1Tad}),
and are given by the poles of $\Gamma(d/2-n_1)$ and $\Gamma(d/2-n_2)$
(with plus sign in front of $d$).

In many cases, calculation of diagrams in the coordinate space
can be simpler than in the momentum space.
In particular, the one-loop propagator diagram of Fig.~\ref{L1}
in the coordinate space is just the product of two propagators.
The massless propagators in the $p$-space and the $x$-space
are related to each other by the Fourier transform:
\begin{align}
&\int \frac{e^{-ipx}}{(-p^2-i0)^n} \frac{d^d p}{(2\pi)^d} =
i 2^{d-2n} \pi^{d/2} \frac{\Gamma(d/2-n)}{\Gamma(n)} \frac{1}{(-x^2+i0)^{d/2-n}}\,,
\label{Foud1}\\
&\int \frac{e^{ipx}}{(-x^2+i0)^n} d^d x =
-i 2^{-2n} \pi^{-d/2} \frac{\Gamma(d/2-n)}{\Gamma(n)} \frac{1}{(-p^2-i0)^{d/2-n}}
\label{Foud2}
\end{align}
(sanity checks: transform $1/(-p^2-i0)^n$ to $x$-space~(\ref{Foud1})
and back~(\ref{Foud2}), and you get $1/(-p^2-i0)^n$;
take the complex conjugate of~(\ref{Foud1}),
rename $x\leftrightarrow p$, $n\leftrightarrow d/2-n$,
and multiply by $(4\pi)^d$, and you get~(\ref{Foud2})).
Multiplying two propagators~(\ref{Foud1}) with the degrees $n_1$ and $n_2$,
we get
\begin{equation*}
- 2^{2(d-n_1-n_2)} \pi^d
\frac{\Gamma(d/2-n_1) \Gamma(d/2-n_2)}{\Gamma(n_1) \Gamma(n_2)}
\frac{1}{(-x^2)^{d-n_1-n_2}}\,.
\end{equation*}
Applying the inverse Fourier transform~(\ref{Foud2}),
we reproduce the result~(\ref{G1}).

The one-loop diagram~(\ref{G1}) is an analytical function
in the complex $p^2$ plane with a cut.
The cut at $p^2>0$, where the real pair production is possible,
begins at the branching point at the threshold $p^2=0$.
This cut comes from the factor $(-p^2)^{-\epsilon}$,
which is equal to $(p^2)^{-\epsilon}\,e^{-i\pi\epsilon}$
at the upper side of the cut and $(p^2)^{-\epsilon}\,e^{i\pi\epsilon}$
at the lower side,
and hence has the discontinuity $-2i(p^2)^{-\epsilon}\sin\pi\epsilon$.
Therefore, the discontinuity of the one-loop diagram~(\ref{G1})
at $\epsilon\to0$ is proportional to the residue of $G(n_1,n_2)$
at $\epsilon=0$.
The discontinuity of the diagram with $n_1=n_2=1$ can be
directly calculated using the Cutkosky rules:
draw a cut across the loop indicating the real intermediate state,
and replace the cut propagators by their discontinuities
\begin{equation}
\frac{1}{p^2+i0} \to -2 \pi i \delta(p^2)\,.
\label{Cut}
\end{equation}

\section{One-loop HQET propagator diagram}
\label{SecHQET1}

\begin{figure}[ht]
\begin{center}
\begin{picture}(32,12)
\put(16,6){\makebox(0,0){\includegraphics{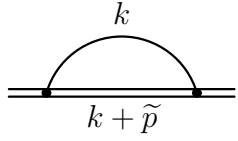}}}
\put(16,-1){\makebox(0,0){$k+\p$}}
\put(16,13){\makebox(0,0){$k$}}
\end{picture}
\end{center}
\caption{One-loop HQET propagator diagram}
\label{H1}
\end{figure}

Now we shall calculate the one-loop HQET propagator diagram
with arbitrary degrees of the denominators (Fig.~\ref{H1})
\begin{equation}
I = \int \frac{d^d k}{(-(k+\p)_0-i0)^{n_1}(-k^2-i0)^{n_2}}\,.
\label{h1}
\end{equation}
It depends only on $\omega=\p_0$, and not on $\vec{p}$.

We cannot use the ordinary Feynman parametrization~(\ref{Feyn}),
because the denominators have different dimensionalities.
Therefore, we make a different change of variables
in the double $\alpha$-parametric integral~(\ref{alpha2}):
$\alpha_1=y\alpha$, $\alpha_2=\alpha$,
and obtain the HQET Feynman parametrization
\begin{equation}
\frac{1}{a_1^{n_1} a_2^{n_2}} = \frac{\Gamma(n_1+n_2)}{\Gamma(n_1) \Gamma(n_2)}
\int_0^\infty \frac{y^{n_1-1}\, dy}
{\left[a_1 y + a_2\right]^{n_1+n_2}}\,.
\label{HQETFeyn}
\end{equation}
If the denominator $a_1$ has dimensionality of energy,
and $a_2$ --- of energy squared, then the Feynman parameter $y$
has dimensionality of energy; it runs from 0 to $\infty$.
Combining the denominators in~(\ref{h1})
and shifting the integration momentum $k\to k-yv/2$, we obtain
\begin{equation}
I = \frac{\Gamma(n_1+n_2)}{\Gamma(n_1) \Gamma(n_2)}
\int \frac{y^{n_1-1}\, dy\, d^d k}
{\left[-k^2+y(y/4-\omega)-i0\right]^{n_1+n_2}}\,.
\label{h1Comb}
\end{equation}
We must have a definite sign of $i0$ in the combined denominator,
therefore, $i0$ terms in both denominators must have the same sign.

We shall assume that $\omega<0$, so that production of a pair
of on-shell particles is not possible.
Using~(\ref{Tadpole}) with $m^2\to y(y/4-\omega)$ and $n\to n_1+n_2$, we obtain
\begin{equation}
I = i \pi^{d/2} \frac{\Gamma(-d/2+n_1+n_2)}{\Gamma(n_1) \Gamma(n_2)}
\int_0^\infty y^{n_1-1}
\left[y\left(\frac{y}{4}-\omega\right)\right]^{d/2-n_1-n_2}\, dy\,.
\label{h1Tad}
\end{equation}
We proceed to the dimensionless variable $z=y/(-\omega)$.
Then the substitution $z/4+1=1/x$ reduces this integral to the Euler B-function:
\begin{equation}
\int_0^\infty y^{d/2-n_2-1} \left(\frac{y}{4}-\omega\right)^{d/2-n_1-n_2}\, dy =
2^{d-2n_2} (-\omega)^{d-n_1-2n_2}
\frac{\Gamma(-d+n_1+2n_2) \Gamma(d/2-n_2)}{\Gamma(-d/2+n_1+n_2)}\,.
\label{ParI}
\end{equation}
Our final result can be written as
\begin{equation}
\begin{split}
& \int \frac{d^d k}{D_1^{n_1} D_2^{n_2}} =
i \pi^{d/2} (-2\omega)^{d-2n_2} I(n_1,n_2)\,,\\
& D_1 = (kv+\omega)/\omega\,, \quad D_2 = -k^2\,,\\
& I(n_1,n_2) =
\frac{\Gamma(-d+n_1+2n_2) \Gamma(d/2-n_2)}{\Gamma(n_1) \Gamma(n_2)}\,.
\end{split}
\label{I1}
\end{equation}
If $n_{1,2}$ are integer, $I(n_1,n_2)$ is proportional to
$I_1=\Gamma(1+2\epsilon)\Gamma(1-\epsilon)$,
the coefficient being a rational function of $d$.

The integral~(\ref{Tadpole}) in $d^d k$ can have UV divergence,
and it produces $\Gamma(-d/2+n_1+n_2)$ in the numerator of~(\ref{h1Tad}).
However, it is too optimistic with respect to UV convergence:
for example, at $n_1=2$, $n_2=1$ it has no pole at $d=4$,
while the original integral~(\ref{h1}) is UV divergent.
The HQET Feynman parameter $y$ has dimensionality of energy
and runs up to $\infty$.
It is natural to expect that this can produce an extra UV divergence.
The region $y\to\infty$ produces $\Gamma(-d+n_1+2n_2)$
in the Feynman parametric integral~(\ref{ParI}).
The ``wrong'' UV $\Gamma$-function is cancelled with the denominator
of~(\ref{ParI}), and $n_1+2n_2$ determines the correct UV behaviour.
The region $y\to0$ produces the IR $\Gamma(d/2-n_2)$.
The rule about negative/positive sign of $d$ in UV/IR $\Gamma$-functions
remains valid.

The one-loop propagator diagram of Fig.~\ref{H1}
in the coordinate space is just the product of two propagators.
The HQET propagators in the $p$-space and the $x$-space
are related to each other by the Fourier transform:
\begin{align}
&\int_{-\infty}^{+\infty} \frac{e^{-i\omega t}}{(-\omega-i0)^n} \frac{d\omega}{2\pi} =
\frac{i^n}{\Gamma(n)} t^{n-1} \theta(t)\,,
\label{Fou11}\\
&\int_0^{+\infty} e^{i\omega t} t^n\, dt =
\frac{(-i)^{n+1} \Gamma(n+1)}{(-\omega-i0)^{n+1}}\,.
\label{Fou12}
\end{align}
The first integral is non-zero only at $t>0$,
when we close the integration contour downwards;
for integer $n$, it follows from the residue at $\omega=-i0$.
In the second integral, we substitute $\omega\to\omega+i0$ for convergence.
Multiplying the HQET propagator~(\ref{Fou11}) with the degree $n_1$
and the massless propagator~(\ref{Foud1}) with the degree $n_2$, we get
\begin{equation*}
- 2^{d-2n_2} \pi^{d/2}
\frac{\Gamma(d/2-n_2)}{\Gamma(n_1) \Gamma(n_2)}
(it)^{n_1+2n_2-d-1} \theta(t)\,.
\end{equation*}
Applying the inverse Fourier transform~(\ref{Fou12}),
we reproduce the result~(\ref{I1}).

The one-loop diagram~(\ref{I1}) is an analytical function
in the complex $\omega$ plane with a cut.
The cut at $\omega>0$, where the real pair production is possible,
begins at the branching point at the threshold $\omega=0$.
This cut comes from the factor $(-\omega)^{-2\epsilon}$,
which has the discontinuity $-2i\omega^{-2\epsilon}\sin2\pi\epsilon$.
Therefore, the discontinuity of the one-loop diagram~(\ref{I1})
at $\epsilon\to0$ is proportional to the residue of $I(n_1,n_2)$
at $\epsilon=0$.
The discontinuity of the diagram with $n_1=n_2=1$ can be
directly calculated using the Cutkosky rules~(\ref{Cut}) and
\begin{equation}
\frac{1}{\p_0+i0} \to -2 \pi i \delta(\p_0)\,.
\label{CutHQET}
\end{equation}

\section{Two-loop massless propagator diagrams}
\label{SecQCD2}

\begin{figure}[ht]
\begin{center}
\begin{picture}(112,20)
\put(56,10){\makebox(0,0){\includegraphics{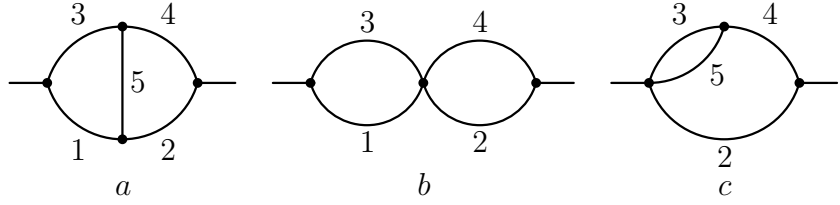}}}
\put(10,1){\makebox(0,0){1}}
\put(22,1){\makebox(0,0){2}}
\put(10,19){\makebox(0,0){3}}
\put(22,19){\makebox(0,0){4}}
\put(18,10){\makebox(0,0){5}}
\put(48.5,2){\makebox(0,0){1}}
\put(63.5,2){\makebox(0,0){2}}
\put(48.5,18){\makebox(0,0){3}}
\put(63.5,18){\makebox(0,0){4}}
\put(96,0){\makebox(0,0){2}}
\put(90,19){\makebox(0,0){3}}
\put(102,19){\makebox(0,0){4}}
\put(95,11){\makebox(0,0){5}}
\put(16,-5){\makebox(0,0)[b]{$a$}}
\put(56,-5){\makebox(0,0)[b]{$b$}}
\put(96,-5){\makebox(0,0)[b]{$c$}}
\end{picture}
\end{center}
\caption{Two-loop massless propagator diagram}
\label{L2}
\end{figure}

There is only one generic topology
of two-loop massless propagator diagrams, Fig.~\ref{L2}a.
If one of the lines is shrunk into a point,
the diagrams of Fig.~\ref{L2}b, c result.
If any two adjacent lines are shrunk into a point,
the diagram contains a no-scale vacuum tadpole,
and hence vanishes.

We write down the diagram of Fig.~\ref{L2}a as
\begin{equation}
\begin{split}
&\int \frac{d^d k_1\, d^d k_2}{D_1^{n_1} D_2^{n_2} D_3^{n_3} D_4^{n_4} D_5^{n_5}} =
- \pi^d (-p^2)^{d-\sum n_i} G(n_1,n_2,n_3,n_4,n_5)\,,\\
&D_1=-(k_1+p)^2\,,\quad D_2=-(k_2+p)^2\,,\\
&D_3=-k_1^2\,,\quad D_4=-k_2^2\,,\quad D_5=-(k_1-k_2)^2\,.
\end{split}
\label{l2}
\end{equation}
It is symmetric with respect to $(1\leftrightarrow2,3\leftrightarrow4)$,
and with respect to $(1\leftrightarrow3,2\leftrightarrow4)$.
If indices of any two adjacent lines are non-positive,
the diagram contains a scale-free vacuum subdiagram,
and hence vanishes.
If $n_5=0$, this is a product of two one-loop diagrams (Fig.~\ref{L2}b):
\begin{equation}
G(n_1,n_2,n_3,n_4,0) = G(n_1,n_3) G(n_2,n_4)\,.
\label{G50}
\end{equation}
If $n_1=0$ (Fig.~\ref{L2}c), the $(3,5)$ integral~(\ref{G1})
gives $G(n_3,n_5)/(-k_2^2)^{n_3+n_5-d/2}$;
this is combined with the denominator 4, and we obtain
\begin{equation}
G(0,n_2,n_3,n_4,n_5) = G(n_3,n_5) G(n_2,n_4+n_3+n_5-d/2)\,.
\label{G10}
\end{equation}
Of course, the cases $n_2=0$, $n_3=0$, $n_4=0$ follow by the symmetry.

When all $n_i>0$, the problem does not immediately reduce
to a repeated use of the one-loop formula~(\ref{G1}).
We shall use a powerful method called integration by parts~\cite{CT}.
It is based on the simple observation that any integral of
$\partial/\partial k_1(\cdots)$
(or $\partial/\partial k_2(\cdots)$)
vanishes (in dimensional regularization no surface terms can appear).
From this, we can obtain recurrence relations which involve
$G(n_1,n_2,n_3,n_4,n_5)$ with different sets of indices.
Applying these relations in a carefully chosen order,
we can reduce any $G(n_1,n_2,n_3,n_4,n_5)$ to trivial ones,
like~(\ref{G50}), (\ref{G10}).

The differential operator $\partial/\partial k_2$
applied to the integrand of~(\ref{l2}) acts as
\begin{equation}
\frac{\partial}{\partial k_2} \to
\frac{n_2}{D_2}2(k_2+p) + \frac{n_4}{D_4}2k_2 + \frac{n_5}{D_5}2(k_2-k_1)\,.
\label{dk1}
\end{equation}
Applying $(\partial/\partial k_2)\cdot k_2$ to the integrand of~(\ref{l2}),
we get a vanishing integral.
On the other hand, from~(\ref{dk1}), $2k_2\cdot k_2=-2D_4$,
$2(k_2+p)\cdot k_2=(-p^2)-D_2-D_4$, $2(k_2-k_1)\cdot k_2=D_3-D_4-D_5$,
we see that this differential operator is equivalent to inserting
\begin{equation*}
d-n_2-n_5-2n_4 + \frac{n_2}{D_2}((-p^2)-D_4) + \frac{n_5}{D_5}(D_3-D_4)
\end{equation*}
under the integral sign (here $(\partial/\partial k_2)\cdot k_2=d$).
Taking into account the definition~(\ref{l2}),
we obtain the recurrence relation
\begin{equation*}
\begin{split}
&(d-n_2-n_5-2n_4)G(n_1,n_2,n_3,n_4,n_5)\\
&\quad{} + n_2 \left[G(n_1,n_2+1,n_3,n_4,n_5)-G(n_1,n_2+1,n_3,n_4-1,n_5)\right]\\
&\quad{} + n_5 \left[G(n_1,n_2,n_3-1,n_4,n_5+1)-G(n_1,n_2,n_3,n_4-1,n_5+1)\right] = 0\,.
\end{split}
\end{equation*}
This relation looks lengthy.
When using the integration-by-parts method,
one has to work with a large number of relations of this kind.
Therefore, special concise notations were invented.
Let's introduce the raising and lowering operators
\begin{equation}
\1\pm G(n_1,n_2,n_3,n_4,n_5) = G(n_1\pm1,n_2,n_3,n_4,n_5)\,,
\label{pm}
\end{equation}
and similar ones for the other indices.
Then our recurrence relation can be written in a shorter
and more easily digestible form
\begin{equation}
\left[d-n_2-n_5-2n_4 + n_2\2+(1-\4-) + n_5\5+(\3--\4-)\right] G = 0\,.
\label{Tria1}
\end{equation}

This is a particular example of the triangle relation.
We differentiate in the loop momentum running along the triangle 254,
and insert the momentum of the line 4 in the numerator.
The differentiation raises the degree of one of the denominators 2, 5, 4.
In the case of the line 4, we get $-2D_4$ in the numerator,
giving just $-2n_4$.
In the case of the line 5, we get the denominator $D_3$
of the line attached to the vertex 45 of our triangle,
minus the denominators $D_4$ and $D_5$.
The case of the line 2 is similar; the denominator of the line
attached to the vertex 24 of our triangle is just $-p^2$,
and it does not influence any index of $G$.
Of course, there are three more relations obtained from~(\ref{Tria1})
by the symmetry.
Another useful triangle relation is derived by applying the operator
$(\partial/\partial k_2)\cdot(k_2-k_1)$:
\begin{equation}
\left[d-n_2-n_4-2n_5 + n_2\2+(\1--\5-) + n_4\4+(\3--\5-)\right] G = 0\,.
\label{Tria2}
\end{equation}
One more is obtained by the symmetry.
Relations of this kind can be written for any diagram having a triangle in it,
when at least two vertices of the triangle each have only a single line
(not belonging to the triangle) attached.

We can obtain a relation from homogeneity of the integral~(\ref{l2}) in $p$.
Applying the operator $p\cdot(\partial/\partial p)$ to the integral~(\ref{l2}),
we see that it is equivalent to the factor $2(d-\sum n_i)$.
On the other hand, explicit differentiation of the integrand gives
$-(n_1/D_1)(-p^2+D_1-D_3)-(n_2/D_2)(-p^2+D_2-D_4)$.
Therefore,
\begin{equation}
\left[ 2(d-n_3-n_4-n_5)-n_1-n_2 + n_1\1+(1-\3-) + n_2\2+(1-\4-) \right] I = 0\,.
\label{homp}
\end{equation}
This is nothing but the sum of the $(\partial/\partial k_2)\cdot k_2$
relation~(\ref{Tria1}) and its mirror-symmetric
$(\partial/\partial k_1)\cdot k_1$ relation.

Another interesting relation is obtained by inserting $(k_1+p)^\mu$
into the integrand of~(\ref{l2}) and taking derivative $\partial/\partial p^\mu$
of the integral.
On the one hand, the vector integral must be proportional to $p^\mu$,
and we can make the substitution
\begin{equation*}
k_1+p \to \frac{(k_1+p)\cdot p}{p^2} p =
\left(1 + \frac{D_1-D_3}{-p^2}\right) \frac{p}{2}
\end{equation*}
in the integrand.
Taking $\partial/\partial p^\mu$ of this vector integral produces~(\ref{l2}) with
\begin{equation*}
\left({\textstyle\frac{3}{2}d-\sum n_i}\right)
\left(1 + \frac{D_1-D_3}{-p^2}\right)
\end{equation*}
inserted into the integrand.
On the other hand, the explicit differentiation in $p$ gives
\begin{equation*}
d + \frac{n_1}{D_1} 2(k_1+p)^2 + \frac{n_2}{D_2} 2(k_2+p)\cdot(k_1+p)\,,\quad
2(k_2+p)\cdot(k_1+p) = D_5-D_1-D_2\,.
\end{equation*}
Therefore, we obtain
\begin{equation}
\left[\tfrac{1}{2}d+n_1-n_3-n_4-n_5
+ \left({\textstyle\frac{3}{2}d-\sum n_i}\right)(\1--\3-)
+ n_2\2+(\1--\5-)\right] G = 0\,.
\label{Larin}
\end{equation}
Three more relations follow from the symmetries.

Expressing $G(n_1,n_2,n_3,n_4,n_5)$ from~(\ref{Tria2}):
\begin{equation}
G(n_1,n_2,n_3,n_4,n_5) =
\frac{n_2\2+(\5--\1-) + n_4\4+(\5--\3-)}{d-n_2-n_4-2n_5} G\,,
\label{Tria}
\end{equation}
we see that the sum $n_1+n_3+n_5$ reduces by 1.
Therefore, applying~(\ref{Tria}) sufficiently many times,
we can reduce an arbitrary $G$ integral with integer indices
to a combination of integrals
with $n_5=0$ (Fig.~\ref{L2}b, (\ref{G50})),
$n_1=0$ (Fig.~\ref{L2}c, (\ref{G10})),
$n_3=0$ (mirror-symmetric to the previous case).
Of course, if $\max(n_2,n_4)<\max(n_1,n_3)$, it may be more efficient
to use the relation mirror-symmetric to~(\ref{Tria2}).
The relation~(\ref{Larin}) also can be used instead of~(\ref{Tria2}).
Thus, any integral $G(n_1,n_2,n_3,n_4,n_5)$ with integer $n_i$
can be expressed as a linear combination of $G_1^2$ and $G_2$,
coefficients being rational functions of $d$.
Here the combinations of $\Gamma$-functions appearing in $n$-loop sunset
massless diagram are
\begin{equation}
G_n = \frac{\Gamma(1+n\epsilon)\Gamma^{n+1}(1-\epsilon)}
{\Gamma(1-(n+1)\epsilon)}\,.
\label{Gn}
\end{equation}

Methods of calculation of three-loop massless propagator diagrams
are considered in~\cite{CT}.

\section{Two-loop HQET propagator diagrams}
\label{SecHQET2}

\begin{figure}[ht]
\begin{center}
\begin{picture}(112,42)
\put(56,21){\makebox(0,0){\includegraphics{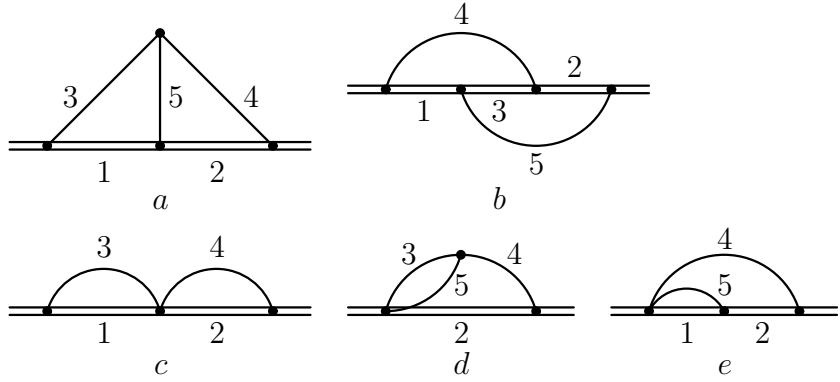}}}
\put(21,16){\makebox(0,0)[b]{$a$}}
\put(13.5,21){\makebox(0,0){1}}
\put(28.5,21){\makebox(0,0){2}}
\put(9,31){\makebox(0,0){3}}
\put(33,31){\makebox(0,0){4}}
\put(23,31){\makebox(0,0){5}}
\put(66,16){\makebox(0,0)[b]{$b$}}
\put(56,29){\makebox(0,0){1}}
\put(66,29){\makebox(0,0){3}}
\put(76,35){\makebox(0,0){2}}
\put(61,42){\makebox(0,0){4}}
\put(71,22){\makebox(0,0){5}}
\put(21,-6){\makebox(0,0)[b]{$c$}}
\put(13.5,-0.5){\makebox(0,0){1}}
\put(28.5,-0.5){\makebox(0,0){2}}
\put(13.5,11){\makebox(0,0){3}}
\put(28.5,11){\makebox(0,0){4}}
\put(61,-6){\makebox(0,0)[b]{$d$}}
\put(61,-0.5){\makebox(0,0){2}}
\put(54,10){\makebox(0,0){3}}
\put(68,10){\makebox(0,0){4}}
\put(61,6){\makebox(0,0){5}}
\put(96,-6){\makebox(0,0)[b]{$e$}}
\put(91,-0.5){\makebox(0,0){1}}
\put(101,-0.5){\makebox(0,0){2}}
\put(96,12){\makebox(0,0){4}}
\put(96,6){\makebox(0,0){5}}
\end{picture}
\end{center}
\caption{Two-loop HQET propagator diagram}
\label{H2}
\end{figure}

There are two generic topologies
of two-loop HQET propagator diagrams, Fig.~\ref{H2}a, b.
If one of the lines is shrunk into a point,
the diagrams of Fig.~\ref{L2}c, d, e result.
If any two adjacent lines are shrunk into a point,
the diagram contains a no-scale vacuum tadpole,
and hence vanishes.

We write down the diagram of Fig.~\ref{H2}a as
\begin{equation}
\begin{split}
&\int \frac{d^d k_1\, d^d k_2}{D_1^{n_1} D_2^{n_2} D_3^{n_3} D_4^{n_4} D_5^{n_5}} =
- \pi^d (-2\omega)^{2(d-n_3-n_4-n_5)} I(n_1,n_2,n_3,n_4,n_5)\,,\\
&D_1 = (k_1+\p)\cdot v/\omega\,,\quad D_2 = (k_2+\p)\cdot v/\omega\,,\\
&D_3 = -k_1^2\,,\quad D_4 = -k_2^2\,,\quad D_5 = -(k_1-k_2)^2\,.
\end{split}
\label{h2}
\end{equation}
It is symmetric with respect to $(1\leftrightarrow3,2\leftrightarrow4)$.
If indices of any two adjacent lines are non-positive,
the diagram contains a scale-free vacuum subdiagram,
and hence vanishes.
If $n_5=0$, this is a product of two one-loop diagrams (Fig.~\ref{H2}c):
\begin{equation}
I(n_1,n_2,n_3,n_4,0) = I(n_1,n_3) I(n_2,n_4)\,.
\label{I50}
\end{equation}
If $n_1=0$ (Fig.~\ref{H2}d), the $(3,5)$ integral~(\ref{G1})
gives $G(n_3,n_5)/(-k_2^2)^{n_3+n_5-d/2}$;
this is combined with the denominator 4, and we obtain
\begin{equation}
I(0,n_2,n_3,n_4,n_5) = G(n_3,n_5) I(n_2,n_4+n_3+n_5-d/2)
\label{I10}
\end{equation}
(and similarly for $n_2=0$).
If $n_3=0$ (Fig.~\ref{H2}e), the $(1,5)$ integral~(\ref{I1})
gives $I(n_1,n_5)\allowbreak/(-2\omega)^{n_1+2n_5-d}$;
this is combined with the denominator 2, and we obtain
\begin{equation}
I(n_1,n_2,0,n_4,n_5) = I(n_1,n_5) I(n_2+n_1+2n_5-d,n_4)
\label{I30}
\end{equation}
(and similarly for $n_4=0$).

When all $n_i>0$, we apply integration by parts~\cite{BG}.
The differential operator $\partial/\partial k_2$
applied to the integrand of~(\ref{h2}) acts as
\begin{equation}
\frac{\partial}{\partial k_2} \to
-\frac{n_2}{D_2}\frac{v}{\omega} + \frac{n_4}{D_4}2k_2 + \frac{n_5}{D_5}2(k_2-k_1)\,.
\label{hdk1}
\end{equation}
Applying $(\partial/\partial k_2)\cdot k_2$ and
$(\partial/\partial k_2)\cdot(k_2-k_1)$ to the integrand of~(\ref{h2}),
we get vanishing integrals.
On the other hand, from~(\ref{hdk1}), $k_2v/\omega=D_2-1$,
$2(k_2-k_1)\cdot k_2=D_3-D_4-D_5$, we see that these differential operators
are equivalent to inserting
\begin{equation*}
\begin{split}
&d-n_2-n_5-2n_4 + \frac{n_2}{D_2} + \frac{n_5}{D_5}(D_3-D_4)\,,\\
&d-n_2-n_4-2n_5 + \frac{n_2}{D_2}D_1 + \frac{n_4}{D_4}(D_3-D_5)
\end{split}
\end{equation*}
under the integral sign.
Therefore, we obtain the triangle relations
\begin{align}
&\left[ d-n_2-n_5-2n_4 + n_2\2+ + n_5\5+(\3--\4-) \right] I = 0\,,
\label{Tri1}\\
&\left[ d-n_2-n_4-2n_5 + n_2\2+\1- + n_4\4+(\3--\5-) \right] I = 0
\label{Tri2}
\end{align}
(two more relations are obtained by $(1\leftrightarrow3,2\leftrightarrow4)$).
Similarly, applying the differential operator
$2\omega(\partial/\partial k_2)\cdot v$ is equivalent to inserting
\begin{equation*}
- 2\frac{n_2}{D_2} + \frac{n_4}{D_4} 4\omega^2 (D_2-1)
+ \frac{n_5}{D_5} 4\omega^2 (D_2-D_1)\,,
\end{equation*}
and we obtain (taking into account the definition~(\ref{h2})
\begin{equation}
\left[ -2n_2\2+ + n_4\4+(\2--1) + n_5\5+(\2--\1-) \right] I = 0
\label{Tri3}
\end{equation}
(there is also the symmetric relation, of course).

We can obtain a relation from homogeneity of the integral~(\ref{h2})
in $\omega$.
Applying the operator $\omega(d/d\omega)$ to the integral~(\ref{h2})
multiplied by $(-\omega)^{-n_1-n_2}$, we see that it is equivalent
to the factor $2(d-n_3-n_4-n_5)-n_1-n_2$.
On the other hand, explicit differentiation of
$(-\omega D_1)^{-n_1}(-\omega D_2)^{-n_2}$ gives $-n_1/D_1-n_2/D_2$.
Therefore,
\begin{equation}
\left[ 2(d-n_3-n_4-n_5)-n_1-n_2 + n_1\1+ + n_2\2+ \right] I = 0\,.
\label{hom}
\end{equation}
This is nothing but the sum of the $(\partial/\partial k_2)\cdot k_2$
relation~(\ref{Tri1}) and its mirror-symmetric
$(\partial/\partial k_1)\cdot k_1$ relation.

When trying to find the most useful combinations of recurrence relations,
it is convenient to manipulate these relations in an algebraic way.
We shall consider the raising and lowering symbols, such as $\1\pm$,
as operators not commuting with $n_1$, etc.
Traditionally, these operators are written to the right of $n_i$ factors.
Of course, any relation may be multiplied (from the left) by any $n_i$ factors.
We can also shift, say, $n_1\to n_1\pm1$ everywhere in the relation.
This operation can be represented by multiplication by $\1\pm$ from the left
followed by commuting this operator to the right, if we assume
\begin{equation}
\1\pm n_1 = (n_1\pm1)\1\pm\,.
\label{comm}
\end{equation}

The most useful combination of recurrence relations
for the integral $I$~(\ref{h2}) is the triangle relation~(\ref{Tri2})
minus $\1-$ times the homogeneity relation~(\ref{hom}):
\begin{equation}
\begin{split}
&\Bigl[ d-n_1-n_2-n_4-2n_5+1\\
&\quad{} - \bigl(2(d-n_3-n_4-n_5)-n_1-n_2+1\bigr)\1-
+ n_4\4+(\3--\5-) \Bigr] I = 0\,.
\end{split}
\label{David}
\end{equation}
Expressing $I(n_1,n_2,n_3,n_4,n_5)$ from this relation:
\begin{equation}
\begin{split}
&I(n_1,n_2,n_3,n_4,n_5) = \frac{1}{d-n_1-n_2-n_4-2n_5+1} \times\\
&\quad\left[(2(d-n_3-n_4-n_5)-n_1-n_2+1)\1- + n_4\4+(\5--\3-)\right] I\,,
\end{split}
\label{David1}
\end{equation}
we see that the sum $n_1+n_3+n_5$ reduces by 1.
Therefore, applying~(\ref{David1}) sufficiently many times,
we can reduce an arbitrary $I$ integral with integer indices
to a combination of integrals
with $n_5=0$ (Fig.~\ref{H2}c, (\ref{I50})),
$n_1=0$ (Fig.~\ref{H2}d, (\ref{I10})),
$n_3=0$ (Fig.~\ref{H2}e, (\ref{I30})).
Of course, if $\max(n_2,n_4)<\max(n_1,n_3)$, it may be more efficient
to use the relation mirror-symmetric to~(\ref{David}).
Thus, any integral $I(n_1,n_2,n_3,n_4,n_5)$ with integer $n_i$
can be expressed as a linear combination of $I_1^2$ and $I_2$,
coefficients being rational functions of $d$.
Here the combinations of $\Gamma$-functions appearing in $n$-loop sunset
HQET diagram are
\begin{equation}
I_n = \Gamma(1+2n\epsilon) \Gamma^n(1-\epsilon)\,.
\label{In}
\end{equation}

We write down the diagram of the second topology, Fig.~\ref{H2}b, as
\begin{equation}
\begin{split}
&\int \frac{d^d k_1\, d^d k_2}{D_1^{n_1} D_2^{n_2} D_3^{n_3} D_4^{n_4} D_5^{n_5}} =
- \pi^d (-2\omega)^{2(d-n_4-n_5)} J(n_1,n_2,n_3,n_4,n_5)\,,\\
&D_1 = (k_1+\p)\cdot v/\omega\,,\quad D_2 = (k_2+\p)\cdot v/\omega\,,\quad
D_3 = (k_1+k_2+\p)\cdot v/\omega\,,\\
&D_4 = -k_1^2\,,\quad D_5 = -k_2^2\,.
\end{split}
\label{j2}
\end{equation}
It is symmetric with respect to $(1\leftrightarrow2,4\leftrightarrow5)$.
If $n_4\le0$, or $n_5\le0$, or two adjacent heavy indices are non-positive,
the diagram vanishes.
If $n_3=0$, this is a product of two one-loop diagrams (Fig.~\ref{H2}c).
If $n_1=0$ or $n_2=0$, this is a diagram of Fig.~\ref{H2}e.
When $n_{1,2,3}$ are all positive, we can insert
\begin{equation*}
1 = D_1 + D_2 - D_3
\end{equation*}
into the integrand of~(\ref{j2}), and obtain
\begin{equation}
J = (\1-+\2--\3-) J\,.
\label{parfrac}
\end{equation}
This reduces $n_1+n_2+n_3$ by 1.
Therefore, applying~(\ref{parfrac}) sufficiently many times,
we can reduce an arbitrary $J$ integral with integer indices
to a combination of integrals with $n_1=0$, $n_2=0$, $n_3=0$.

Methods of calculation of three-loop HQET propagator diagrams
are considered in~\cite{G}.

\section{Renormalization of QCD}
\label{SecQCDr}

The QCD Lagrangian expressed via the bare (unrenormalized) quantities
(denoted by the subscript 0) is
\begin{equation}
L = \sum_i \overline{q}_{i0} (i\D_0-m_{i0}) q_{i0}
- \frac{1}{4} G^a_{0\mu\nu} G^{a\mu\nu}_0
- \frac{1}{2a_0} \left(\partial_\mu A_0^\mu\right)^2
+ (\partial_\mu \overline{c}_0^a) (D_0^\mu c_0^a)\,,
\label{Lagr0}
\end{equation}
where $D_{0\mu}q_0=(\partial_\mu-i g_0 A_{0\mu}^a t^a)q_0$,
$a_0$ is the gauge fixing parameter, $c$ is the ghost field,
and $D_{0\mu}c_0^a=(\partial_\mu\delta^{ab}-g_0 f^{abc}A_\mu^c)c_0^b$.
The renormalized quantities are related to the bare ones by
\begin{equation}
q_{i0} = Z_q^{1/2} q_i\,, \quad
A_0 = Z_A^{1/2} A\,, \quad
c_0 = Z_c^{1/2} c\,, \quad
g_0 = Z_\alpha^{1/2} g\,, \quad
m_{i0} = Z_m m_i\,, \quad
a_0 = Z_A a\,,
\label{renorm}
\end{equation}
where renormalization factors have the minimal structure
\begin{equation}
Z = 1 + \frac{Z_{11}}{\epsilon} \frac{\alpha_s}{4\pi}
+ \left(\frac{Z_{22}}{\epsilon^2}+\frac{Z_{21}}{\epsilon}\right)
\left(\frac{\alpha_s}{4\pi}\right)^2 + \cdots
\label{minim}
\end{equation}

The Lagrangian has dimensionality $[L]=d$,
because the action $S = \int L d^d x$
is exactly dimensionless in the space-time with any $d$.
The gluon kinetic term, which is the $g_0^0$ term
in $\frac{1}{4} G^a_{0\mu\nu} G^{a\mu\nu}_0$,
has the structure $(\partial A_0)^2$;
hence, the dimensionality of the gluon field is $[A_0]=1-\epsilon$.
Similarly, from the quark kinetic terms, the dimensionality
of the quark fields is $[q_{i0}]=\frac{3}{2}-\epsilon$.
The covariant derivative $D_{0\mu}=\partial_\mu-i g_0 A_{0\mu}^a t^a$
has dimensionality 1, hence the dimensionality of the coupling constant
is $[g_0]=\epsilon$.

We define $\alpha_s$ to be exactly dimensionless:
\begin{equation}
\alpha_s = \overline{\mu}\,^{-2\epsilon} \frac{g^2}{4\pi}\,, \quad
\overline{\mu}\,^2 = \mu^2 \frac{e^\gamma}{4\pi}\,, \quad
\text{or} \quad
\frac{\alpha_s}{4\pi} = \mu^{-2\epsilon} Z_\alpha^{-1}
\frac{g_0^2}{(4\pi)^{d/2}} e^{-\gamma\epsilon}\,.
\label{alphas}
\end{equation}
Here $\mu$ is the renormalization scale,
and its rescaling $\overline{\mu}$
(introduced in the \MS{} renormalization scheme)
allows one to get rid of the Euler constant $\gamma$ and $\log 4\pi$
in all equations in the limit $\epsilon\to0$
when they are written via $\alpha_s$.
The renormalized QCD coupling $\alpha_s$ satisfies
the renormalization group equation
\begin{equation}
\frac{d\log\alpha_s}{d\log\mu} = -2\epsilon-2\beta(\alpha_s)\,, \quad
\beta(\alpha_s) = \frac{1}{2} \frac{d\log Z_\alpha}{d\log\mu}
= \beta_0 \frac{\alpha_s}{4\pi}
+ \beta_1 \left(\frac{\alpha_s}{4\pi}\right)^2 + \cdots
\label{rengroup}
\end{equation}
where
\begin{equation}
\beta_0 = \frac{11}{3} C_A - \frac{4}{3} T_F n_f\,,
\label{beta}
\end{equation}
and $\beta_{1,2,3}$ are also known.
Here $T_F=\frac{1}{2}$, $n_f$ is the number of flavours.
All the bare quantities, including $g_0$, are $\mu$-independent.
Substituting the expression for $\alpha_s$ via $g_0^2$ and $Z_\alpha$
into~(\ref{rengroup}), we find
\begin{equation}
Z_\alpha = 1 - \beta_0 \frac{\alpha_s}{4\pi\epsilon}
+ \left(\beta_0^2-\tfrac{1}{2}\beta_1\epsilon\right)
\left(\frac{\alpha_s}{4\pi\epsilon}\right)^2 + \cdots
\label{Zalpha}
\end{equation}
Hence, $Z_{22}$ contains no new information:
it can be expressed via the one-loop term.

\begin{figure}[ht]
\begin{center}
\begin{picture}(111,9)
\put(55.5,4.5){\makebox(0,0){\includegraphics{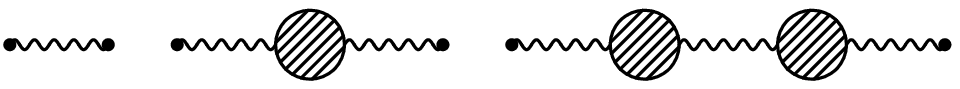}}}
\put(14.5,4.5){\makebox(0,0){+}}
\put(48.5,4.5){\makebox(0,0){+}}
\put(99.5,4.5){\makebox(0,0){+}}
\put(103,4.5){\makebox(0,0)[l]{$\cdots$}}
\end{picture}
\end{center}
\caption{Structure of diagrams for the gluon propagator}
\label{GlueProp}
\end{figure}

The bare (unrenormalized) gluon propagator $-iD_{0\mu\nu}(p)$
has the structure (Fig.~\ref{GlueProp})
\begin{equation}
\begin{split}
-iD_{0\mu\nu}(p) =& -iD^0_{\mu\nu}(p)
+ (-i) D^0_{\mu\alpha}(p) i\Pi^{\alpha\beta}(p) (-i) D^0_{\beta\nu}(p)\\
&{} + (-i) D^0_{\mu\alpha}(p) i\Pi^{\alpha\beta}(p) (-i) D^0_{\beta\gamma}(p)
i\Pi^{\gamma\delta}(p) (-i) D^0_{\gamma\nu}(p) + \cdots
\end{split}
\label{Glue}
\end{equation}
where
\begin{equation}
D^0_{\mu\nu}(p) = \frac{1}{p^2} \left( g_{\mu\nu}
- (1-a_0) \frac{p_\mu p_\nu}{p^2} \right)
\label{Glue0}
\end{equation}
is the free gluon propagator,
and the gluon self-energy (polarization operator) $i\Pi_{\mu\nu}(p)$
is the sum of one-particle-irreducible gluon self-energy diagrams
(which cannot be separated into two disconnected parts
by cutting a single gluon line).
The series~(\ref{Glue}) implies the equation
\begin{equation}
D_{0\mu\nu}(p) = D^0_{\mu\nu}(p)
+ D^0_{\mu\alpha}(p) \Pi^{\alpha\beta}(p) D_{0\beta\nu}(p)\,.
\label{GlueDyson}
\end{equation}
To solve this equation, it is convenient to introduce the inverse tensor
$A^{-1}_{\mu\nu}$ of a symmetric tensor $A^{\mu\nu}$
satisfying $A^{-1}_{\mu\alpha}A^{\alpha\nu}=\delta_\mu^\nu$.
If
\begin{equation*}
A_{\mu\nu} = A_\bot \left(g_{\mu\nu}-\frac{p_\mu p_\nu}{p^2}\right)
+ A_{||} \frac{p_\mu p_\nu}{p^2}\,,
\end{equation*}
then
\begin{equation*}
A^{-1}_{\mu\nu} = A^{-1}_\bot \left(g_{\mu\nu}-\frac{p_\mu p_\nu}{p^2}\right)
+ A^{-1}_{||} \frac{p_\mu p_\nu}{p^2}\,.
\end{equation*}
Using these notations, the equation~(\ref{GlueDyson}) can be rewritten as
\begin{equation}
D^{-1}_{0\mu\nu}(p) = (D^0)^{-1}_{\mu\nu}(p) - \Pi_{\mu\nu}(p)\,.
\label{GlueDyson2}
\end{equation}
Due to the Ward identity $\Pi_{\mu\nu}(p)p^\nu=0$,
the gluon self-energy is transverse
\begin{equation}
\Pi_{\mu\nu}(p) = (p^2 g_{\mu\nu} - p_\mu p_\nu) \Pi(p^2)\,.
\label{GlueTrans}
\end{equation}
Therefore, the gluon propagator is
\begin{equation}
D_{0\mu\nu}(p) =
\frac{1}{p^2(1-\Pi(p^2))} \left(g_{\mu\nu}-\frac{p_\mu p_\nu}{p^2}\right)
+ a_0 \frac{p_\mu p_\nu}{(p^2)^2}\,.
\label{Glue1}
\end{equation}
Its longitudinal part gets no corrections.
The renormalized gluon propagator is related to the bare one by
$D_{0\mu\nu}(p) = Z_A D_{\mu\nu}(p)$.
The gluon field renormalization constant $Z_A$ is constructed to make
the transverse part of $D_{\mu\nu}$ finite in the limit $\epsilon\to0$:
\begin{equation}
D_{\mu\nu}(p) = D_\bot(p^2) \left(g_{\mu\nu}-\frac{p_\mu p_\nu}{p^2}\right)
+ a \frac{p_\mu p_\nu}{(p^2)^2}\,.
\label{Dren}
\end{equation}
At the same time, it converts $a_0$ to the renormalized gauge parameter $a$.
This is the reason why the gauge parameter is renormalized~(\ref{renorm})
by the same constant $Z_A$.

\begin{figure}[ht]
\begin{center}
\begin{picture}(102,17)
\put(51,8.5){\makebox(0,0){\includegraphics{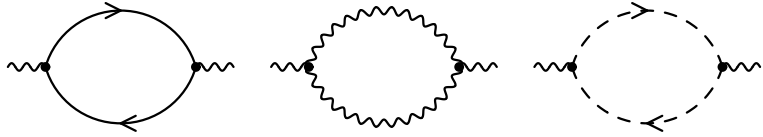}}}
\end{picture}
\end{center}
\caption{One-loop gluon self-energy}
\label{Glue1Loop}
\end{figure}

In the one-loop approximation (Fig.~\ref{Glue1Loop}),
we can use the methods of Sect.~\ref{SecQCD1} to obtain
\begin{align}
&\Pi(p^2) = \frac{\Pi_\mu^\mu(p)}{(d-1)p^2} = \frac{g_0^2(-p^2)^{-\epsilon}}{(4\pi)^{d/2}}
\frac{(d-2)G_1}{(d-1)(d-3)(d-4)}4P + \cdots\,,
\label{Pi1}\\
&P = T_F n_f
- \frac{3d-2 + (d-1)(2d-7)(1-a_0) - \tfrac{1}{4}(d-1)(d-4)(1-a_0)^2}{4(d-2)} C_A
\nonumber
\end{align}
where $G_1$ is defined in~(\ref{Gn}).
Writing $G_{0\mu\nu}(p) = Z_A G_{\mu\nu}(p)$,
where the gluon field renormalization constant
has the minimal form~(\ref{minim}),
and requiring that the renormalized gluon propagator $D_{\mu\nu}(p)$
is finite in the limit $\epsilon\to0$,
we can find $Z_A$ in the one-loop approximation.

It is more convenient to present results for anomalous dimensions
instead of renormalization constants, because anomalous dimensions
contain the same information but are more compact.
An anomalous dimension is defined by
\begin{equation}
\gamma(\alpha_s) = \frac{d\log Z}{d\log\mu} = \gamma_0 \frac{\alpha_s}{4\pi}
+ \gamma_1 \left(\frac{\alpha_s}{4\pi}\right)^2 + \cdots
\label{ADim}
\end{equation}
If $Z$ is gauge invariant, then, differentiating~(\ref{minim})
and using~(\ref{rengroup}), we find $Z_{11}=-\frac{1}{2}\gamma_0$,
$Z_{22}=\frac{1}{8}\gamma_0(\gamma_0+2\beta_0)$,
$Z_{21}=-\frac{1}{4}\gamma_1$.
If $\gamma_0$ linearly depends on $a$
(as is the case for $Z_A$, for example),
then $Z_{22}$ contains an additional term
from the $\mu$-dependence of $a$~(\ref{renorm})
containing the one-loop anomalous dimension of the gluon field:
\begin{equation}
Z = 1 - \frac{1}{2} \gamma_0 \frac{\alpha_s}{4\pi\epsilon}
+ \frac{1}{8} \left[ \gamma_0(\gamma_0+2\beta_0)
+ \gamma_{A0} \frac{d\gamma_0}{da} a - 2 \gamma_1 \epsilon \right]
\left(\frac{\alpha_s}{4\pi\epsilon}\right)^2 + \cdots
\label{Z2}
\end{equation}
Hence, $Z_{22}$ contains no new information:
it can be expressed via the one-loop term.
For the anomalous dimension of the gluon field itself,
we obtain from~(\ref{Pi1})
\begin{equation}
\gamma_A = \left[ \left(a-\tfrac{13}{3}\right) C_A
+ \tfrac{8}{3} T_F n_f \right] \frac{\alpha_s}{4\pi} + \cdots
\label{gammaA}
\end{equation}

\begin{figure}[ht]
\begin{center}
\begin{picture}(111,9)
\put(55.5,4.5){\makebox(0,0){\includegraphics{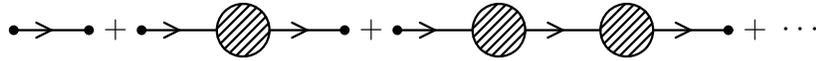}}}
\put(14.5,4.5){\makebox(0,0){+}}
\put(48.5,4.5){\makebox(0,0){+}}
\put(99.5,4.5){\makebox(0,0){+}}
\put(103,4.5){\makebox(0,0)[l]{$\cdots$}}
\end{picture}
\end{center}
\caption{Structure of diagrams for the quark propagator}
\label{QuarkProp}
\end{figure}

The bare (unrenormalized) quark propagator $iS_0(p)$
has the structure (Fig.~\ref{QuarkProp})
\begin{equation}
iS_0(p) = iS^0(p)
+ iS^0(p) (-i)\Sigma(p) iS^0(p)
+ iS^0(p) (-i)\Sigma(p) iS^0(p) (-i)\Sigma(p) iS^0(p) + \cdots
\label{Quark}
\end{equation}
where
\begin{equation}
S^0(p) = \frac{1}{\rlap/p-m_0} = \frac{\rlap/p+m_0}{p^2-m^2}
\label{Quark0}
\end{equation}
is the free quark propagator,
and the quark self-energy (mass operator) $-i\Sigma(p)$
is the sum of one-particle-irreducible quark self-energy diagrams
(which cannot be separated into two disconnected parts
by cutting a single quark line).
The series~(\ref{Quark}) implies the equation
\begin{equation}
S_0(p) = S^0(p) + S^0(p) \Sigma(p) S_0(p)
\label{QuarkDyson}
\end{equation}
with the solution
\begin{equation}
S_0(p) = \frac{1}{(S^0)^{-1}(p)-\Sigma(p)} = \frac{1}{\rlap/p-m_0-\Sigma(p)}\,.
\label{Quark1}
\end{equation}
The quark self-energy has the structure
\begin{equation}
\Sigma(p) = \rlap/p \Sigma_V(p^2) + m_0 \Sigma_S(p^2)\,.
\label{Sig}
\end{equation}
Therefore, the quark propagator is
\begin{equation}
S_0(p) = \frac{1}{1-\Sigma_V(p^2)}
\frac{1}{\rlap/p - (1-\Sigma_V(p^2))^{-1}(1+\Sigma_S(p^2))m_0}
= Z_q S(p)\,,
\label{Quark2}
\end{equation}
and the renormalization constants $Z_q$, $Z_m$
are constructed to make $Z_q (1-\Sigma_V)$ and $Z_m Z_q (1+\Sigma_S)$ finite.

\begin{figure}[ht]
\begin{center}
\begin{picture}(32,9.5)
\put(16,4.75){\makebox(0,0){\includegraphics{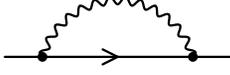}}}
\end{picture}
\end{center}
\caption{One-loop quark self-energy}
\label{Quark1Loop}
\end{figure}

For a massless quark in the one-loop approximation (Fig.~\ref{Quark1Loop}),
\begin{equation}
\Sigma_{V}(p^2) = C_F \frac{g_0^2(-p^2)^{-\epsilon}}{(4\pi)^{d/2}}
\frac{d-2}{(d-3)(d-4)} G_1 a_0\,.
\label{Si1}
\end{equation}
Extracting the $1/\epsilon$ pole, we find the anomalous dimension
\begin{equation}
\gamma_q = 2 a C_F \frac{\alpha_s}{4\pi} + \cdots
\label{gammaq}
\end{equation}
UV divergences don't depend on masses,
therefore, this result is also valid for a massive quark.
Note that there is no UV divergence in the Landau gauge $a=0$.

\section{Renormalization of HQET}
\label{SecHQETr}

The HQET Lagrangian expressed via the bare (unrenormalized) quantities is
\begin{equation}
L = \overline{\Q}_0 iv\cdot D_0 \Q_0 + L_{\text{QCD}}\,, \quad
\Q_0 = \Z_Q^{1/2} \Q\,.
\label{LagrHQET0}
\end{equation}
All the renormalization constants in~(\ref{renorm}) are the same as in QCD,
where the heavy flavour $Q$ is not counted in $n_f$.
In order to find the new constant $\Z_Q$,
we need to calculate the heavy quark propagator in HQET.

\begin{figure}[ht]
\begin{center}
\begin{picture}(111,9)
\put(55.5,4.5){\makebox(0,0){\includegraphics{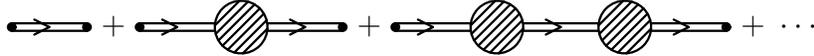}}}
\put(14.5,4.5){\makebox(0,0){+}}
\put(48.5,4.5){\makebox(0,0){+}}
\put(99.5,4.5){\makebox(0,0){+}}
\put(103,4.5){\makebox(0,0)[l]{$\cdots$}}
\end{picture}
\end{center}
\caption{Structure of diagrams for the heavy quark propagator in HQET}
\label{HQETProp}
\end{figure}

The bare (unrenormalized) static quark propagator $i\S_0(\omega)$
has the structure (Fig.~\ref{HQETProp})
\begin{equation}
i\S_0(\omega) = i\S^0(\omega)
+ i\S^0(\omega) (-i)\Si(\omega) i\S^0(\omega)
+ i\S^0(\omega) (-i)\Si(\omega) i\S^0(\omega) (-i)\Si(\omega) i\S^0(\omega) + \cdots
\label{HQuark}
\end{equation}
where $\S^0(\omega) = 1/\omega$ is the free HQET propagator,
and the static quark self-energy (mass operator) $-i\Si(\omega)$
is the sum of one-particle-irreducible HQET self-energy diagrams
(which cannot be separated into two disconnected parts
by cutting a single heavy quark line).
Summing this series, we obtain
\begin{equation}
\S_0(\omega) = \frac{1}{\omega-\Si(\omega)}\,.
\label{HQuark1}
\end{equation}

\begin{figure}[ht]
\begin{center}
\begin{picture}(32,9.5)
\put(16,4.75){\makebox(0,0){\includegraphics{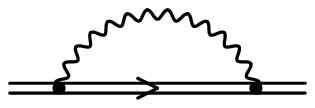}}}
\end{picture}
\end{center}
\caption{One-loop heavy quark self-energy in HQET}
\label{HQET1Loop}
\end{figure}

In the one-loop approximation (Fig.~\ref{HQET1Loop}),
\begin{equation*}
\Si(\omega) = i C_F \int \frac{d^d k}{(2\pi)^d}
i g_0 v^\mu \frac{i}{kv+\omega} i g_0 v^\nu \frac{-i}{k^2}
\left( g_{\mu\nu} - (1-a_0) \frac{k_\mu k_\nu}{k^2} \right)
\end{equation*}
After contraction over the indices,
the second term in the brackets contains $(kv)^2=(kv+\omega-\omega)^2$.
This factor can be replaced by $\omega^2$, because all integrals
without $kv+\omega$ in the denominator are scale-free and hence vanish.
Using the definition~(\ref{I1}), we get
\begin{equation*}
\Si(\omega) = - C_F \frac{g_0^2(-2\omega)^{1-2\epsilon}}{(4\pi)^{d/2}}
\left[ 2 I(1,1) + \tfrac{1}{2} (1-a_0) I(1,2) \right]\,,
\end{equation*}
and, finally,
\begin{equation}
\Si(\omega) = C_F \frac{g_0^2(-2\omega)^{1-2\epsilon}}{(4\pi)^{d/2}}
\frac{I_1}{d-4} A\,, \quad
A = a_0-1-\frac{2}{d-3}\,,
\label{HQETSi1}
\end{equation}
where $I_1$ is defined in~(\ref{In}).

Therefore, with the one-loop accuracy, the heavy quark propagator in HQET is
\begin{equation}
\omega \S_0(\omega) = 1 + C_F \frac{g_0^2(-2\omega)^{-2\epsilon}}{(4\pi)^{d/2}}
\frac{I_1}{d-4} 2A\,.
\label{HQETS1}
\end{equation}
In the coordinate space,
\begin{equation}
\S_0(t) = \S^0(t) \left[ 1 + C_F \frac{g_0^2 (it/2)^{2\epsilon}}{(4\pi)^{d/2}}
\Gamma(-\epsilon) A + \cdots \right]\,, \quad
\S^0(t) = - i \theta(t)\,.
\label{HQETS1c}
\end{equation}
Now we re-express $\S_0$ via the renormalized quantities $\alpha_s$, $a$
(this is trivial at the present accuracy),
and find $\Z_Q$ of the form~(\ref{minim}) from the requirement that
the renormalized propagator $\S(\omega)=\Z_Q^{-1}\S_0(\omega)$ is finite
in the limit $\epsilon\to0$.
The result for the anomalous dimension is
\begin{equation}
\ga_Q = 2 (a-3) C_F \frac{\alpha_s}{4\pi} + \cdots
\label{gammaQ1}
\end{equation}
Note that there is no UV divergence in the Yennie gauge $a=3$.

\begin{figure}[ht]
\begin{center}
\begin{picture}(69,34.5)
\put(34.5,17.25){\makebox(0,0){\includegraphics{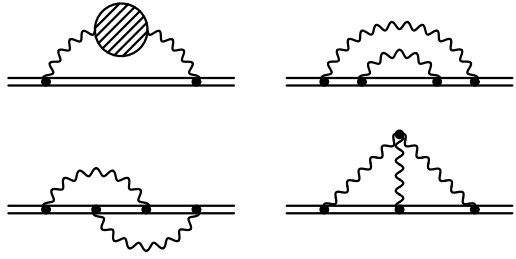}}}
\end{picture}
\end{center}
\caption{Two-loop heavy quark self-energy in HQET}
\label{HQET2Loop}
\end{figure}

Two-loop diagrams for the heavy-quark self-energy
are shown in Fig.~\ref{HQET2Loop}.
The first one contains the one-loop gluon self-energy~(\ref{Pi1});
it can be easily calculated using~(\ref{I1}),
and is proportional to $I_2$~(\ref{In}):
\begin{equation}
\Si_{2a}(\omega) = - C_F \frac{g_0^4 (-2\omega)^{1-4\epsilon}}{(4\pi)^d}
\frac{(d-2) I_2}{(d-3) (d-4)^2 (d-6) (2d-7)} 4P\,.
\label{HQETSi2a}
\end{equation}
The second one is also recursively one-loop;
it is proportional to $C_F^2 I_2$.
It is also not difficult to understand its dependence
on the gauge parameter $a_0$.
The exterior loop contains the heavy-quark line with a non-integer index $n$
(in fact, $n=1+2\epsilon$).
The propagator~(\ref{Glue0}) of the exterior gluon produces,
as compared to the Feynman gauge $a_0=1$, the extra factor
\begin{equation*}
1 + \frac{1-a_0}{4} \frac{I(n,2)-2I(n-1,2)+I(n-2,2)}{I(n,1)} =
1 + \tfrac{1}{2} (1-a_0) (d-3) = - \tfrac{1}{2} (d-3) A\,.
\end{equation*}
The same is true for the interior gluon line,
with $n=1$, as in the one-loop case~(\ref{HQETSi1}).
Therefore, the result contains $A^2$:
\begin{equation}
\Si_{2b}(\omega) = - C_F^2 \frac{g_0^4 (-2\omega)^{1-4\epsilon}}{(4\pi)^d}
\frac{(d-3) I_2}{(d-4)^2 (2d-7)} A^2\,.
\label{HQETSi2b}
\end{equation}

\begin{figure}[p]
\begin{center}
\begin{picture}(80,12)
\put(40,6){\makebox(0,0){\includegraphics{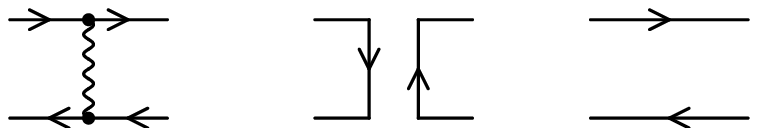}}}
\put(24.5,6){\makebox(0,0){$\displaystyle{}=\frac{1}{2}\Biggl[\Biggr.$}}
\put(54,6){\makebox(0,0){$\displaystyle{}-\frac{1}{N_c}$}}
\put(79,6){\makebox(0,0){$\displaystyle\Biggl.\Biggr]$}}
\end{picture}
\end{center}
\caption{Cvitanovi\'c algorithm}
\label{Cvit}
\end{figure}

\begin{figure}[p]
\begin{center}
\begin{picture}(105,42)
\put(52.5,21){\makebox(0,0){\includegraphics{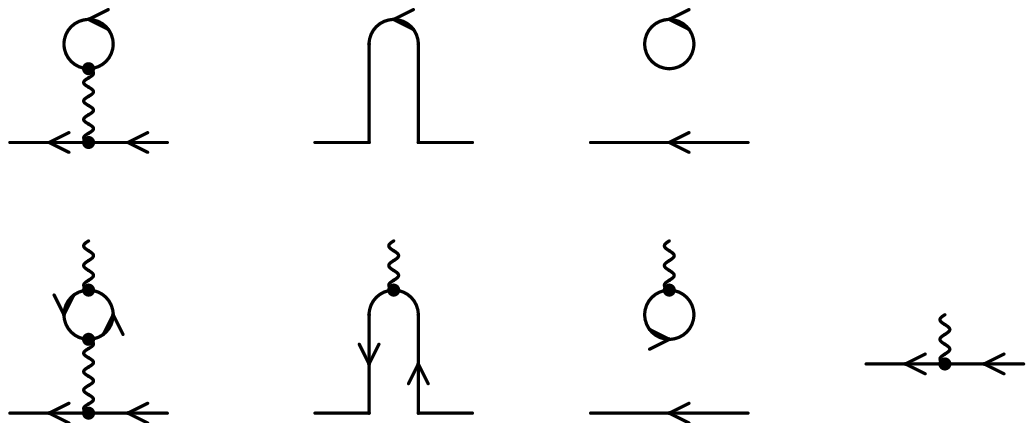}}}
\put(24.5,6){\makebox(0,0){$\displaystyle{}=\frac{1}{2}\Biggl[\Biggr.$}}
\put(54,6){\makebox(0,0){$\displaystyle{}-\frac{1}{N_c}$}}
\put(82,6){\makebox(0,0){$\displaystyle\Biggl.\Biggr]=\frac{1}{2}$}}
\put(24.5,33.5){\makebox(0,0){$\displaystyle{}=\frac{1}{2}\Biggl[\Biggr.$}}
\put(54,33.5){\makebox(0,0){$\displaystyle{}-\frac{1}{N_c}$}}
\put(82,33.5){\makebox(0,0){$\displaystyle\Biggl.\Biggr]=0$}}
\end{picture}
\end{center}
\caption{Checking the coefficients}
\label{Cvit2}
\end{figure}

\begin{figure}[p]
\begin{center}
\begin{picture}(148,34)
\put(74,17){\makebox(0,0){\includegraphics{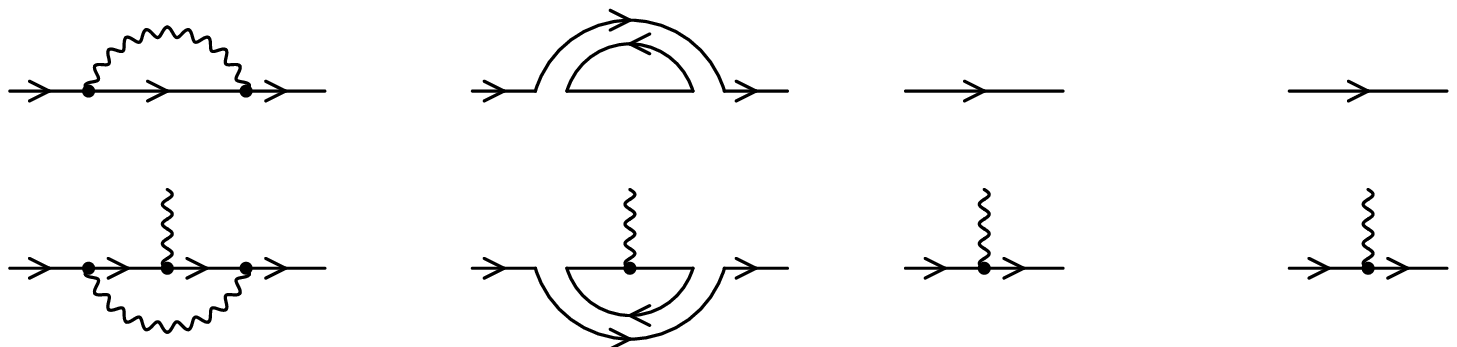}}}
\put(40.5,8){\makebox(0,0){$\displaystyle{}=\frac{1}{2}\Biggl[\Biggr.$}}
\put(86,8){\makebox(0,0){$\displaystyle{}-\frac{1}{N_c}$}}
\put(119.5,8){\makebox(0,0){$\displaystyle\Biggl.\Biggr]=-\frac{1}{2N_c}$}}
\put(40.5,26){\makebox(0,0){$\displaystyle{}=\frac{1}{2}\Biggl[\Biggr.$}}
\put(86,26){\makebox(0,0){$\displaystyle{}-\frac{1}{N_c}$}}
\put(119.5,26){\makebox(0,0){$\displaystyle\Biggl.\Biggr]=\frac{N_c^2-1}{2N_c}$}}
\end{picture}
\end{center}
\caption{Sample applications of Cvitanovi\'c algorithm}
\label{Cvit3}
\end{figure}

\begin{figure}[p]
\begin{center}
\begin{picture}(112,12)
\put(56,6){\makebox(0,0){\includegraphics{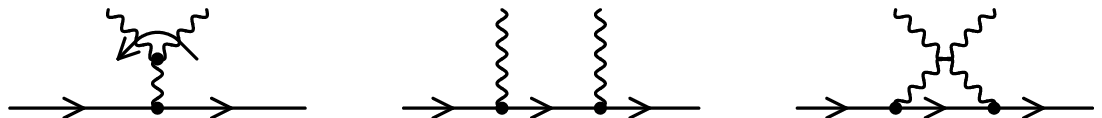}}}
\put(36,6){\makebox(0,0){$=$}}
\put(76,6){\makebox(0,0){$-$}}
\end{picture}
\end{center}
\caption{Three-gluon vertex}
\label{Cvit4}
\end{figure}

The colour factor of the third diagram can be easily found
using the Cvitanovi\'c algorithm~\cite{Cv} (Fig.~\ref{Cvit}).
The gluon exchange between two quark lines is replaced
by exchange by a quark-antiquark pair,
minus a colourless exchange which compensates its colour-singlet part.
Correctness of the coefficients in this identity can be checked
by closing the upper quark line, and closing it with attaching a gluon (Fig.~\ref{Cvit2}).
Two example applications of this algorithms are shown in Fig.~\ref{Cvit3}:
the first one is a calculation of $C_F$,
and the second one shows that the colour factor of the diagram Fig.~\ref{HQET2Loop}c is
$C_F\left(-\frac{1}{2N_c}\right)=C_F\left(C_F-\frac{1}{2}C_A\right)$.
The colour factor of the three-gluon vertex $if^{abc}$
is defined as via commutator $[t^a,t^b]=if^{abc}t^c$ (Fig.~\ref{Cvit4}).
Therefore, the colour factor of the diagram Fig.~\ref{HQET2Loop}d
is the difference of those of Fig.~\ref{HQET2Loop}b and Fig.~\ref{HQET2Loop}c,
$\frac{1}{2} C_F C_A$.

The diagram Fig.~\ref{HQET2Loop}c,
after killing one of the heavy-quark lines~(\ref{parfrac}),
yields $I_1^2$ or $I_2$;
it contains $A^2$ for the same reason as Fig.~\ref{HQET2Loop}b:
\begin{equation}
\Si_{2c}(\omega) = C_F \left(C_F - \tfrac{1}{2}C_A\right)
\frac{g_0^4 (-2\omega)^{1-4\epsilon}}{(4\pi)^d}
\left[\frac{2I_1^2}{(d-4)^2} - \frac{I_2}{(d-4)(2d-7)}\right] A^2\,.
\label{HQETSi2c}
\end{equation}
The diagram Fig.~\ref{HQET2Loop}d vanishes in the Feynman gauge,
because the three-gluon vertex vanishes after contraction
with three identical vectors $v$.
It also vanishes, if longitudinal parts of all three gluon propagators
are taken, and hence contains no $(1-a_0)^3$ term.
The result is
\begin{equation}
\Si_{2d}(\omega) = - C_F C_A \frac{g_0^4 (-2\omega)^{1-4\epsilon}}{(4\pi)^d}
\frac{1}{(d-4)^2} \left[ I_1^2 + \frac{I_2}{2(2d-7)} \right] A (1-a_0)\,.
\label{HQETSi2d}
\end{equation}

Collecting these results together,
we obtain the bare heavy-quark propagator with the two-loop accuracy
\begin{align}
&\omega \S_0(\omega) = 1
+ C_F \frac{g_0^2(-2\omega)^{-2\epsilon}}{(4\pi)^{d/2}} \frac{I_1}{d-4} 2A
\label{HQETS2}\\
&{} + C_F \frac{g_0^4(-2\omega)^{-4\epsilon}}{(4\pi)^d} \frac{1}{(d-4)^2}
\Biggl[ \frac{8(d-2)}{(d-3)(d-6)(2d-7)} T_F n_f I_2
+ 2 A^2 C_F I_2 - \frac{4A}{d-3} C_A I_1^2
\nonumber\\
& + \frac{2}{(d-3)^2(d-6)} \left( \frac{(d-2)^2 (d-5)}{(d-3)(2d-7)}
+ (d^2-4d+5)A - \tfrac{1}{4}(d^2-9d+16)(d-3)A^2 \right) C_A I_2 \Biggr]\,.
\nonumber
\end{align}
Now we re-express it via $\alpha_s$~(\ref{alphas}) and $a$~(\ref{renorm});
in the one-loop term, the $\alpha_s$ corrections
in $Z_\alpha$~(\ref{Zalpha}), (\ref{beta})
and $Z_A$~(\ref{Z2}), (\ref{gammaA}) are necessary.
After that, we expand in $\epsilon$, and find the renormalization constant
$\Z_Q$ of the minimal form~(\ref{minim}) such that $\S(\omega)=\Z_Q^{-1} \S_0(\omega)$
is finite at $\epsilon\to0$.
It must have the form~(\ref{Z2}).
We arrive at the anomalous dimension
\begin{equation}
\ga_Q = 2 (a-3) C_F \frac{\alpha_s}{4\pi}
+ C_F \left( \frac{3a^2+24a-179}{6} C_A + \frac{32}{3} T_F n_f \right)
\left(\frac{\alpha_s}{4\pi}\right)^2 + \cdots
\label{gammaQ2}
\end{equation}
Its difference with the QCD quark field anomalous dimension~(\ref{gammaq})
is gauge invariant up to two loops:
\begin{equation}
\ga_Q - \gamma_q = - 6 C_F \frac{\alpha_s}{4\pi}
- C_F \left( \frac{127}{3} C_A - 3 C_F - \frac{44}{3} T_F n_f \right)
\left(\frac{\alpha_s}{4\pi}\right)^2 + \cdots\,.
\label{gammaQq}
\end{equation}

\section{Heavy Electron Effective Theory}
\label{SecHEET}

Now we make a short digression into the abelian version of HQET --- the
Heavy Electron Effective Theory, an effective field theory of QED
describing interaction of a single electron with soft photons.
It is obtained by setting $C_F\to1$, $C_A\to0$,
$g_0\to e_0$, $\alpha_s\to\alpha$.
This theory was considered long ago,
and is called the Bloch--Nordsieck model.

\begin{figure}[ht]
\begin{center}
\begin{picture}(158,61)
\put(79,30.5){\makebox(0,0){\includegraphics{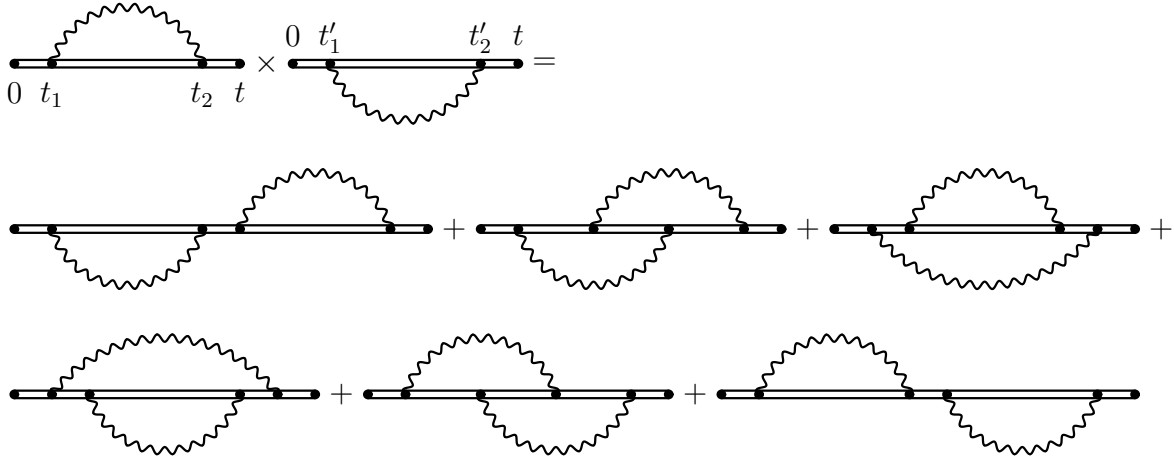}}}
\put(1,50){\makebox(0,0)[t]{0}}
\put(6,50){\makebox(0,0)[t]{$t_1$}}
\put(26,50){\makebox(0,0)[t]{$t_2$}}
\put(31,50){\makebox(0,0)[t]{$t$}}
\put(38,57.5){\makebox(0,0)[t]{0}}
\put(43,58){\makebox(0,0)[t]{$t_1'$}}
\put(63,58){\makebox(0,0)[t]{$t_2'$}}
\put(68,57.5){\makebox(0,0)[t]{$t$}}
\put(34.5,52.5){\makebox(0,0){$\times$}}
\put(71.5,52.5){\makebox(0,0){=}}
\put(59.5,30.5){\makebox(0,0){+}}
\put(106.5,30.5){\makebox(0,0){+}}
\put(153.5,30.5){\makebox(0,0){+}}
\put(44.5,8.5){\makebox(0,0){+}}
\put(91.5,8.5){\makebox(0,0){+}}
\end{picture}
\end{center}
\caption{Exponentiation theorem}
\label{Exp}
\end{figure}

Suppose we calculate the one-loop correction to the heavy electron propagator
in the coordinate space.
Let's multiply this correction by itself (Fig.~\ref{Exp}).
We get an integral in $t_1$, $t_2$, $t_1'$, $t_2'$
with $0<t_1<t_2<t$, $0<t_1'<t_2'<t$.
Ordering of primed and non-primed integration times can be arbitrary.
The integration area is subdivided into 6 regions,
corresponding to 6 diagrams in Fig.~\ref{Exp}.
This is twice the two-loop correction to the propagator.
Continuing this drawing exercise,
we see that the one-loop correction cubed
is $3!$ times the three-loop correction, and so on.
Therefore, the exact all-order propagator
is the exponent of the one-loop term:
\begin{equation}
\S_0(t) = \S^0(t) \exp \left[ \frac{e_0^2 (it/2)^{2\epsilon}}{(4\pi)^{d/2}}
\Gamma(-\epsilon) A \right]\,.
\label{HEETS}
\end{equation}

\begin{figure}[p]
\begin{center}
\begin{picture}(86,9.5)
\put(43,4.75){\makebox(0,0){\includegraphics{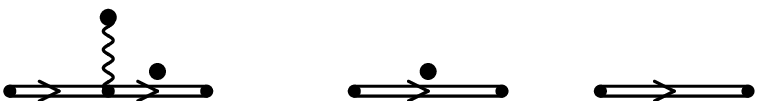}}}
\put(28.5,1){\makebox(0,0){${}=e_0\Biggl[\Biggr.$}}
\put(56,1){\makebox(0,0){$-$}}
\put(80,1){\makebox(0,0){$\Biggl.\Biggr]$}}
\end{picture}
\end{center}
\caption{Ward identity for the free electron propagator}
\label{WardS}
\end{figure}

In this theory, $Z_A=1$, because there exist no loops
which can be inserted into the photon propagator.
Now we are going to show that $Z_\alpha=1$, too.
To this end, let's consider the sum of all one-particle-irreducible
vertex diagrams, not including the external leg propagators ---
the electron-photon proper vertex.
It has the same structure as the tree-level term: $i e_0 v^\mu \Ga$, $\Ga=1+\La$,
where $\La$ is the sum of all unrenormalized diagrams
starting from one loop.
Let's multiply the vertex by the incoming photon momentum $q_\mu$.
This product can be simplified by the Ward identity
for the electron propagator (Fig.~\ref{WardS}):
\begin{equation}
i \S^0(\p') \; i e_0 v\cdot q \; i \S^0(\p) =
i e_0 \frac{i}{\p'\cdot v} \left[ \p'\cdot v - \p\cdot v \right]
\frac{i}{\p\cdot v} = 
i e_0 \left[ \S^0(\p') - \S^0(\p) \right]\,.
\label{wardS}
\end{equation}
In the Figure, the fat dot at the end of the photon line
means multiplication by $q_\mu$;
a dot near an electron propagator means that its momentum
is shifted by $q$.

\begin{figure}[p]
\begin{center}
\begin{picture}(138,131)
\put(69,65.5){\makebox(0,0){\includegraphics{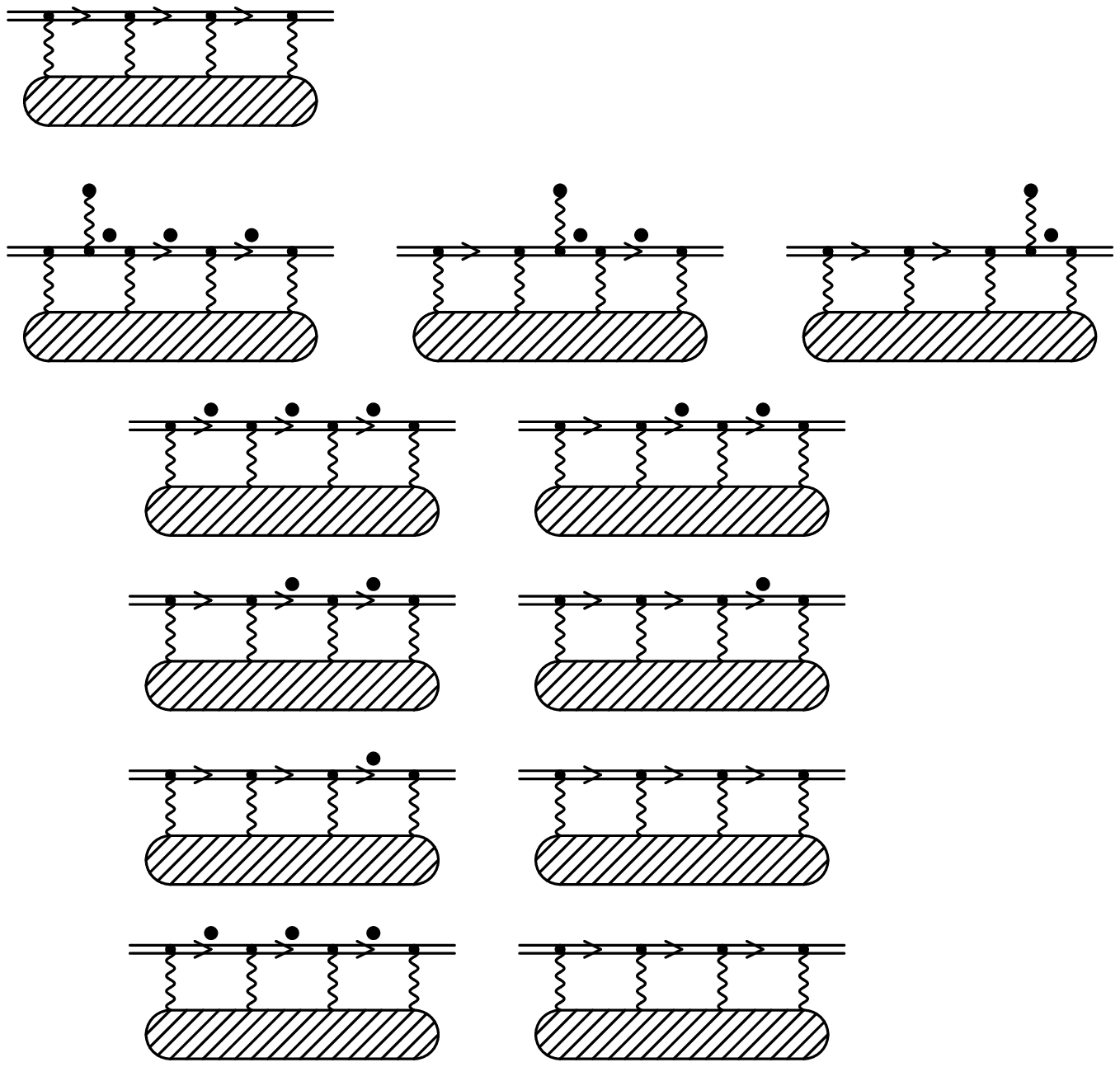}}}
\put(46,129.25){\makebox(0,0){$\Longrightarrow$}}
\put(45,100.5){\makebox(0,0){$+$}}
\put(93,100.5){\makebox(0,0){$+$}}
\put(6,79){\makebox(0,0){${}=e_0\Biggl[\Biggr.$}}
\put(60,79){\makebox(0,0){$-$}}
\put(12,57.5){\makebox(0,0){$+$}}
\put(60,57.5){\makebox(0,0){$-$}}
\put(12,36){\makebox(0,0){$+$}}
\put(60,36){\makebox(0,0){$-$}}
\put(110,36){\makebox(0,0){$\Biggl.\Biggr]$}}
\put(6,14.5){\makebox(0,0){${}=e_0\Biggl[\Biggr.$}}
\put(60,14.5){\makebox(0,0){$-$}}
\put(110,14.5){\makebox(0,0){$\Biggl.\Biggr]$}}
\end{picture}
\end{center}
\caption{Ward identity for the electron-photon vertex}
\label{WardV}
\end{figure}

Starting from each diagram for $\Si$, we can obtain a set of diagrams for $\La$
by inserting the external photon vertex into each electron propagator.
After multiplying by $q_\mu$, each diagram in this set becomes a difference.
All terms cancel each other, except the extreme ones (Fig.~\ref{WardV}),
and we obtain the Ward identity
\begin{equation}
\La(\omega,\omega') = - \frac{\Si(\omega')-\Si(\omega)}{\omega'-\omega}
\quad\text{or}\quad
\Ga(\omega',\omega) =
\frac{\S_0^{-1}(\omega')-\S_0^{-1}(\omega)}{\omega'-\omega}\,.
\label{wardV}
\end{equation}
The vertex function is thus also known to all orders.
The charge renormalization constant $Z_\alpha$ is obtained from the requirement
that the renormalized vertex function $g_0\Ga Z_A^{1/2}\Z_Q$ is finite.
The factor $\Z_Q$ transforms $\S_0^{-1}$ in~(\ref{wardV}) into $\S^{-1}$
and hence makes $\Ga$ finite.
Therefore, the remaining factor $(Z_\alpha Z_A)^{1/2}=1$
(this is also true in QED).
In the Bloch--Nordsieck model, $Z_A=1$ and hence $Z_\alpha=1$.

Due to the absence of charge and photon field renormalization,
we may replace $e_0\to e$, $a_0\to a$ in the bare propagator~(\ref{HEETS}).
It is made finite by the minimal (in the sense of~(\ref{minim}))
renormalization constant, which is just the exponent of the one-loop term
\begin{equation}
\Z_Q = \exp \left[ -(a-3)\frac{\alpha}{4\pi\epsilon} \right]\,,
\end{equation}
and the anomalous dimension is exactly equal to the one-loop contribution
\begin{equation}
\ga_Q = 2 (a-3) \frac{\alpha}{4\pi}\,.
\label{gammaAQ1}
\end{equation}
Note that the electron propagator is finite to all orders in the Yennie gauge.

What information useful for the real HQET can be extracted from this
simple abelian model?
First of all, we can obtain the $C_F^2$ term in~(\ref{HQETS2})
by the Fourier transformation, without explicit calculation.
There should be no $C_F^2$ term
in the two-loop field anomalous dimension~(\ref{gammaQ2}).
The Ward identity for the gluon propagator is more involved than Fig.~\ref{WardS}:
in addition to the difference of propagators, it contains ghost terms.
The Ward identity for the quark-gluon vertex in HQET contains an extra term
as compared to~(\ref{wardV}).
At one loop, it comes from the diagram with three-gluon vertex,
and is proportional to $C_A$.
Therefore, $Z_\alpha Z_A$ is no longer unity.

\section{$1/m$ corrections to HQET Lagrangian}
\label{Sec1m}

As discussed in Sec.~\ref{SecHQET},
we are interested in processes
with characteristic momenta and energies $\omega\ll m$.
The heavy quark effective theory is constructed to reproduce
QCD $S$-matrix elements expanded up to some order in $\omega/m$.
There is a ``folk theorem'' that any $S$-matrix
having all the required properties follows from some Lagrangian.
Therefore, the HQET Lagrangian is constructed as a series in $1/m$
containing all operators having the necessary symmetries,
with arbitrary coefficients.
These coefficients are tuned to reproduce several QCD $S$-matrix elements
expanded to some degree of $\omega/m$.
We should perform this matching for a sufficient number of amplitudes
to fix all the coefficients in the Lagrangian.
After that, we can use this HQET Lagrangian instead of the QCD one
for calculating other amplitudes.

The HQET Lagrangian is not unique,
because the heavy quark field $\Q$ can be redefined.
Such field transformations can be used to eliminate
all time derivatives $D_0$ acting on $\Q$,
except in the leading term~(\ref{HQETlagr})~\cite{KT}.

The velocity $v$ can be varied by a small amount $\delta v\lesssim\omega/m$
without violating applicability of HQET or changing its predictions.
This reparametrization invariance~\cite{LM}
relates terms of different orders in $1/m$.

At the level of $1/m$ terms, the heavy quark spin symmetry
and the superflavour symmetry are violated by interaction
of the heavy quark chromomagnetic moment
with the magnetic component of the gluon field.
This leads to hyperfine splittings between states
which were degenerate in the infinite mass limit (such as $B$ and $B^*$),
as well as to violation of leading-order relations among form factors.
First, we are going to discuss the simpler case of a scalar heavy quark.
We shall return to the realistic spin $\frac{1}{2}$ case
at the end of this Section.

The only dimension-5 operator in scalar HQET
not containing $D_0$ acting on $\Q$ is $\Q^+\vec{D}\,^2\Q$.
Therefore, the Lagrangian is
\begin{equation}
L = \Q^+ iD_0 \Q + \frac{C_k}{2m} \Q^+ \vec{D}\,^2 \Q + \cdots
\label{KinLagr}
\end{equation}
The additional term is the heavy quark kinetic energy.
This Lagrangian leads to the dispersion law of a free quark
$\p_0=p_0-m=C_k\frac{\vec{p}\,^2}{2m}$.
Therefore, at tree level, $C_k=1$.
The Lagrangian~(\ref{KinLagr}) can be rewritten in a covariant form:
\begin{equation}
L_v = \Q^+_v iv\cdot D \Q_v - \frac{C_k}{2m} \Q^+_v D_\bot^2 \Q_v + \cdots
\label{CKinLagr}
\end{equation}
where $D_\bot=D-v(v\cdot D)$.
More accurately, this term should be written as $\frac{C_k^0}{2m}\widetilde{O}_k^0$,
where $\widetilde{O}_k^0=-\Q^+_{v0} D_{\bot0}^2 \Q_{v0}$
is the bare kinetic energy operator.
It is related to the renormalized one as $\widetilde{O}_k^0=\Z_k(\mu)\widetilde{O}_k(\mu)$,
and hence the term in the Lagrangian is $\frac{C_k(\mu)}{2m}\widetilde{O}_k(\mu)$,
where $C_k(\mu)=\Z_k(\mu)C_k^0$.

\begin{figure}[ht]
\begin{center}
\begin{picture}(72,10)
\put(36,5){\makebox(0,0){\includegraphics{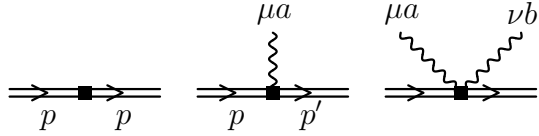}}}
\put(6,-4){\makebox(0,0)[b]{$p$}}
\put(16,-4){\makebox(0,0)[b]{$p$}}
\put(31,-4){\makebox(0,0)[b]{$p$}}
\put(41,-4){\makebox(0,0)[b]{$p'$}}
\put(36,10){\makebox(0,0)[b]{$\mu a$}}
\put(53,10){\makebox(0,0)[b]{$\mu a$}}
\put(69,10){\makebox(0,0)[b]{$\nu b$}}
\end{picture}
\end{center}
\caption{Kinetic-energy vertices}
\label{KinVert}
\end{figure}

The kinetic energy term gives the new vertices (Fig.~\ref{KinVert}):
$i\frac{C_k^0}{2m}\p_\bot^2$,
$i\frac{C_k^0}{2m}g_0t^a(p+p')_\bot^\mu$,
$i\frac{C_k^0}{2m}g_0^2(t^a t^b+t^b t^a)g_\bot^{\mu\nu}$
(where $g_\bot^{\mu\nu}=g^{\mu\nu}-v^\mu v^\nu$).

In Sect.~\ref{SecHQETr}, we considered the sum of all
one-particle-irreducible heavy-quark self energy diagrams $-i\Si(\omega)$
in the infinite mass limit.
Let's denote $-i\frac{C_k^0}{2m}\Si_k(\omega,p_\bot^2)$ the sum of all bare
one-particle-irreducible self energy diagrams at the order $1/m$.
Each of these diagrams contains a single kinetic-energy vertex (Fig.~\ref{KinVert}).
Let's consider the variation of $\Si$ at $v\to v+\delta v$
for an infinitesimal $\delta v$ ($v\cdot\delta v=0$).
There are two sources of this variation.
Expansion of heavy-quark propagators $1/(\p\cdot v+i0)$
produces insertions $i\p_i\cdot\delta v$ into each propagator in turn.
Variations of quark-gluon vertices produce $ig_0t^a\delta v^\mu$
for each vertex in turn.
Now let's consider the variation of $\Si_k$ at $p_\bot\to p_\bot+\delta p_\bot$
for an infinitesimal $\delta p_\bot$.
No-gluon kinetic vertices (Fig.~\ref{KinVert}) produce
$i\frac{C_k^0}{m}\p_i\cdot\delta p_\bot$;
single-gluon kinetic vertices (Fig.~\ref{KinVert}) produce
$i\frac{C_k^0}{m}g_0t^a\delta p_\bot^\mu$;
two-gluon kinetic vertices do not change.
Therefore,
\begin{equation}
\frac{\partial\Si_k}{\partial p_\bot^\mu} =
2 \frac{\partial\Si}{\partial v^\mu}\,.
\label{Repar1}
\end{equation}
This is the Ward identity of reparametrization invariance.
Taking into account $\partial\Si_k/\partial p_\bot^\mu=
2(\partial\Si_k/\partial p_\bot^2)p_\bot^\mu$ and
$\partial\Si/\partial v^\mu=(d\Si/d\omega)p_\bot^\mu$,
we obtain
\begin{equation}
\frac{\partial\Si_k}{\partial p_\bot^2} = \frac{d\Si}{d\omega}\,.
\label{Repar2}
\end{equation}
The right-hand side does not depend on $p_\bot^2$, and hence
\begin{equation}
\Si_k(\omega,p_\bot^2) = \frac{d\Si(\omega)}{d\omega} p_\bot^2 + \Si_{k0}(\omega)\,.
\label{Repar3}
\end{equation}

This result can also be understood in a more direct way.
The momentum $p_\bot$ flows through the heavy quark line.
No-gluon kinetic vertices are quadratic in it;
one-gluon vertices are linear;
two-gluon vertices are $p_\bot$ independent.
The $p_\bot^2$ term comes from diagrams with a no-gluon kinetic vertex.
Terms linear in $p_\bot$ vanish due to the rotational symmetry.
The coefficient of $p_\bot^2$ in a no-gluon kinetic vertex is $i\frac{C_k}{2m}$.
Therefore, the coefficient of $p_\bot^2$ in the sum of all diagrams
is the sum of the leading-order HQET diagrams with a unit operator insertion
into each heavy-quark propagator in turn.
This sum is just $-i\frac{d\Si}{d\omega}$,
and hence we arrive at~(\ref{Repar3}) again.

As we discussed in the beginning of this Section,
coefficients in the HQET Lagrangian are obtained by equating
on-shell scattering amplitudes in full QCD and in HQET
with the required accuracy in $1/m$.
The prerequisite of this matching is the requirement
that the mass shell itself is the same in both theories,
with the accuracy considered.
The mass shell is defined as position of the pole of the full quark propagator.
In QCD it is $p_0=\sqrt{m^2+\vec{p}\,^2}$,
where $m$ is the on-shell mass (see Sect.~\ref{SecQCDos} for more details).
To the first order in $1/m$, this means $\omega=\frac{\vec{p}\,^2}{2m}$.
In HQET, the mass shell is zero of the denominator of the bare heavy-quark propagator
\begin{equation}
\S_0(p) = \frac{1}{\omega - \Si(\omega) - \frac{C_k^0}{2m}
\left[ \vec{p}\,^2 - \frac{d\Si(\omega)}{d\omega} \vec{p}\,^2 + \Si_{k0}(\omega) \right]}\,.
\label{HQETmshell}
\end{equation}
In Sect.~\ref{SecHQETos} we shall obtain at the two-loop level
$\Si(0)=0$ and $\Si_{k0}(0)=0$.
I don't know if this is true in higher orders or not.
If not, these equalities can be restored by adding a residual mass counterterm
to the Lagrangian, this does not contradict to any general requirements.
The mass shell is
\begin{equation}
\left[ 1 - \frac{d\Si(\omega)}{d\omega} \right]_{\omega=0} \omega =
\frac{C_k^0}{2m} \left[ 1 - \frac{d\Si(\omega)}{d\omega} \right]_{\omega=0} \vec{p}\,^2\,.
\label{HQETmshell2}
\end{equation}
It is correct if $C_k^0=\Z_k^{-1}(\mu)C_k(\mu)=1$.
The minimal~(\ref{minim}) renormalization constant $\Z_k$ has to make $C_k(\mu)$ finite;
here this means $\Z_k=1$.
The kinetic energy operator is not renormalized;
its anomalous dimension is zero in all orders.
The coefficient of the kinetic energy operator in the HQET Lagrangian is exactly unity,
\begin{equation}
C_k(\mu) = 1\,.
\label{ReparK}
\end{equation}
to all orders in perturbation theory, due to the reparametrization invariance!

\begin{figure}[ht]
\begin{center}
\begin{picture}(47,10)
\put(23.5,5){\makebox(0,0){\includegraphics{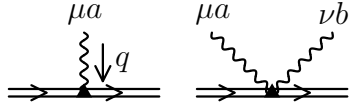}}}
\put(15,5){\makebox(0,0)[l]{$q$}}
\put(11,10){\makebox(0,0)[b]{$\mu a$}}
\put(28,10){\makebox(0,0)[b]{$\mu a$}}
\put(44,10){\makebox(0,0)[b]{$\nu b$}}
\end{picture}
\end{center}
\caption{Chromomagnetic-interaction vertices}
\label{MagVert}
\end{figure}

In the case of a spin $\frac{1}{2}$ heavy quark,
there is one more dimension 5 operator, in addition to the kinetic energy ---
chromomagnetic interaction~\cite{EH2,FGL}
\begin{equation}
L = \Q^+ iD_0 \Q + \frac{C_k}{2m} \Q^+ \vec{D}\,^2 \Q
- \frac{C_m}{2m} \Q^+ \vec{B}\cdot\vec{\sigma} \Q + \cdots
\label{MagLagr}
\end{equation}
or, in covariant notations,
\begin{equation}
L_v = \overline{\Q}_v iv\cdot D \Q_v - \frac{C_k}{2m} \overline{\Q}_v D_\bot^2 \Q_v
+ \frac{C_m}{4m} \overline{\Q}_v G_{\mu\nu} \sigma^{\mu\nu} \Q_v + \cdots
\label{MagLagrC}
\end{equation}
where $G_{\mu\nu}=gG^a_{\mu\nu}t^a$
and $\sigma^{\mu\nu}=\frac{i}{2}[\gamma^\mu,\gamma^\nu]$.
In the $v$ rest frame, only chromomagnetic components of $G_{\mu\nu}$
contribute, because $\sigma^{0i}$ sandwiched between $\overline{\Q}$ and $\Q$
yields zero.
Again, it is more accurate to write this term as
$C_m^0\widetilde{O}_m^0=C_m(\mu)\widetilde{O}_m(\mu)$,
where $\widetilde{O}_m^0=\Z_m(\mu)\widetilde{O}_m(\mu)$
is the bare chromomagnetic operator, and $C_m^0=\Z_m^{-1}(\mu)C_m(\mu)$.
The kinetic term in~(\ref{MagLagr}) does not violate the heavy quark spin symmetry
(because it contains no spin matrix between $\Q^+$ and $\Q$),
while the chromomagnetic term violates it, producing hyperfine splittings.
The chromomagnetic interaction gives the new vertices (Fig.~\ref{MagVert}):
$\frac{C_m^0}{2m}gt^a\sigma^{\nu\mu}q_\nu$,
$\frac{C_m^0}{2m}g^2[t^a,t^b]\sigma^{\mu\nu}$.
The arguments leading to~(\ref{ReparK}) remain valid in the spin $\frac{1}{2}$ case.
The coefficient of the chromomagnetic interaction $C_m(\mu)$ is not related
to the lower order term in $1/m$ by the reparametrization invariance,
and can only be calculated by QCD/HQET matching (Sect.~\ref{SecCMag}).

\section{On-shell renormalization of QCD}
\label{SecQCDos}

\begin{figure}[ht]
\begin{center}
\begin{picture}(32,12)
\put(16,6){\makebox(0,0){\includegraphics{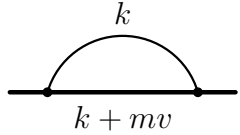}}}
\put(16,-1){\makebox(0,0){$k+mv$}}
\put(16,13){\makebox(0,0){$k$}}
\end{picture}
\end{center}
\caption{One-loop on-shell propagator diagram}
\label{OS1}
\end{figure}

Now we are going to discuss calculation of on-shell propagator diagrams
of a massive quark in QCD.
Let's write the one-loop diagram with arbitrary degrees
of denominators (Fig.~\ref{OS1}) as
\begin{equation}
\int \frac{d^d k}{\left[m^2-(k+mv)^2-i0\right]^{n_1}(-k^2-i0)^{n_2}} =
i\pi^{d/2} m^{d-2(n_1+n_2)} M(n_1,n_2)\,.
\label{os1}
\end{equation}
After the Wick rotation $k_0=ik_{E0}$, $k^2=-k_E^2$
and transformation to the dimensionless integration momentum $K=k_E/m$,
it becomes
\begin{equation}
\int \frac{d^d K}{(K^2-2iK_0)^{n_1}(K^2)^{n_2}} =
\pi^{d/2} M(n_1,n_2)\,.
\label{os1a}
\end{equation}
Now let's compare it with the one-loop HQET diagram propagator $I(n_1,n_2)$~(\ref{I1}).
In terms of the dimensionless Euclidean integration momentum $K=k_E/(-2\omega)$,
it has the form
\begin{equation}
\int \frac{d^d K}{(1-2iK_0)^{n_1}(K^2)^{n_2}} =
\pi^{d/2} I(n_1,n_2)\,.
\label{hqet1}
\end{equation}
The on-shell integral~(\ref{os1a}) can be cast into the HQET form~(\ref{hqet1})
using inversion~\cite{BG2} $K=K'/K^{\prime2}$.
The on-shell denominator
$K^2-2iK_0=\left(1-2iK'_0\right)/K^{\prime2}$
produces the HQET denominator.
The integration measure becomes $d^dK=K^{d-1}dKd\Omega=
\left(K'\right)^{-d-1}dK'd\Omega=d^dK'/\left(K^{\prime2}\right)^d$.
The final result is
\begin{equation}
M(n_1,n_2) = I(n_1,d-n_1-n_2) =
\frac{\Gamma(d-n_1-2n_2)\Gamma(-d/2+n_1+n_2)}{\Gamma(n_1)\Gamma(d-n_1-n_2)}\,.
\label{M1}
\end{equation}
This result can also be obtained using the Feynman parametrization~(\ref{Feyn}).
If $n_{1,2}$ are integer, $M(n_1,n_2)$ is proportional to $M_1=\Gamma(1+\epsilon)$,
the coefficient being a rational function of $d$.

The inversion interchanges UV and IR behaviour of an integral.
Therefore, the poles of $\Gamma(d-n_1-2n_2)$ are IR divergences
(sometimes called on-shell divergences in this case),
and the poles of $\Gamma(-d/2+n_1+n_2)$ are UV divergences.
The usual rule about the sign of $d$ in $\Gamma$ functions applies.

\begin{figure}[ht]
\begin{center}
\begin{picture}(87,48)
\put(43.5,24){\makebox(0,0){\includegraphics{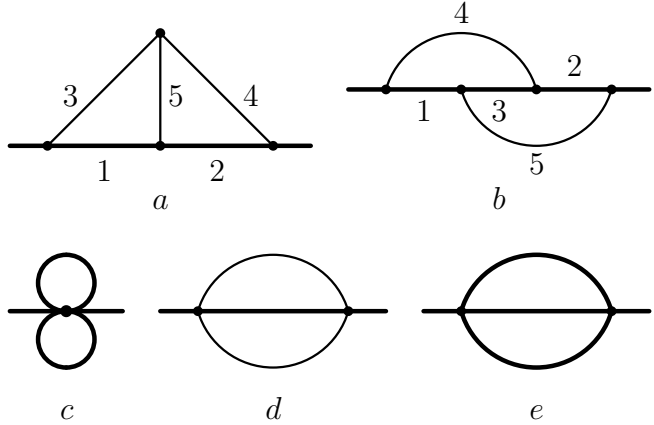}}}
\put(21,23){\makebox(0,0)[b]{$a$}}
\put(13.5,28){\makebox(0,0){1}}
\put(28.5,28){\makebox(0,0){2}}
\put(9,38){\makebox(0,0){3}}
\put(33,38){\makebox(0,0){4}}
\put(23,38){\makebox(0,0){5}}
\put(66,23){\makebox(0,0)[b]{$b$}}
\put(56,36){\makebox(0,0){1}}
\put(66,36){\makebox(0,0){3}}
\put(76,42){\makebox(0,0){2}}
\put(61,49){\makebox(0,0){4}}
\put(71,29){\makebox(0,0){5}}
\put(8.5,-5){\makebox(0,0)[b]{$c$}}
\put(36,-5){\makebox(0,0)[b]{$d$}}
\put(71,-5){\makebox(0,0)[b]{$e$}}
\end{picture}
\end{center}
\caption{Two-loop on-shell propagator diagrams}
\label{OS2}
\end{figure}

There are two two generic topologies of two-loop on-shell
propagator diagrams, Fig.~\ref{OS2}a, b.
We shall start from the simpler type $M$, Fig.~\ref{OS2}a:
\begin{equation}
\begin{split}
&\int \frac{d^d k_1\,d^d k_2}{D_1^{n_1}D_2^{n_2}D_3^{n_3}D_4^{n_4}D_5^{n_5}} =
- \pi^d m^{2(d-\sum n_i)} M(n_1,n_2,n_3,n_4,n_5)\,,\\
&D_1 = m^2-(k_1+mv)^2\,,\quad
D_2 = m^2-(k_2+mv)^2\,,\\
&D_3 = -k_1^2\,,\quad
D_4 = -k_2^2\,,\quad
D_5 = -(k_1-k_2)^2\,.
\end{split}
\label{os2}
\end{equation}
Using inversion, we can relate it to the HQET two-loop propagator diagram.
The on-shell denominators $D_{1,2}$ produce the HQET denominators,
just as in the one-loop case.
The denominator $D_5$ becomes $(K_1-K_2)^2=(K_1'-K_2')^2/(K_1^{\prime2}K_2^{\prime2})$.
We obtain~\cite{BG2}
\begin{equation}
M(n_1,n_2,n_3,n_4,n_5) = I(n_1,n_2,d-n_1-n_3-n_5,d-n_2-n_4-n_5,n_5)\,.
\label{M2}
\end{equation}
However, this relation is not particularly useful for calculation
of $M(n_1,n_2,n_3,n_4,n_5)$,
because this HQET integral contains two non-integer indices.
The integrals $M$ can be calculated
using integration by parts recurrence relations.
This is not so easy as in the massless case (Sect.~\ref{SecQCD2})
and the HQET case (Sect.~\ref{SecHQET2}).
We shall not discuss the algorithm here;
it can be found in the original literature~\cite{B,FT}.
The conclusion is that any integral $M(n_1,n_2,n_3,n_4,n_5)$
with integer indices $n_i$ can be expressed as a linear combination
of $M_1^2$ (Fig.~\ref{OS2}c) and $M_2$ (Fig.~\ref{OS2}d),
coefficients being rational functions of $d$.
Here the combinations of $\Gamma$-functions appearing in $n$-loop
on-shell sunset diagrams with a single massive line are
\begin{equation}
M_n = \frac{\Gamma(1+(n-1)\epsilon)\Gamma(1+n\epsilon)\Gamma(1-2n\epsilon)
\Gamma^n(1-\epsilon)}{\Gamma(1-n\epsilon)\Gamma(1-(n+1)\epsilon)}\,.
\label{Mn}
\end{equation}

The type $N$ (Fig.~\ref{OS2}b)
\begin{equation}
\begin{split}
&\int \frac{d^d k_1\,d^d k_2}{D_1^{n_1}D_2^{n_2}D_3^{n_3}D_4^{n_4}D_5^{n_5}} =
- \pi^d m^{2(d-\sum n_i)} N(n_1,n_2,n_3,n_4,n_5)\,,\\
&D_1 = m^2-(k_1+mv)^2\,,\quad
D_2 = m^2-(k_2+mv)^2\,,\quad
D_3 = m^2-(k_1+k_2+mv)^2\,,\\
&D_4 = -k_1^2\,,\quad
D_4 = -k_2^2
\end{split}
\label{os2n}
\end{equation}
is more difficult.
The integrals $N$ can be expressed~\cite{B,FT}, using integration by parts,
as linear combinations of $M_1^2$ (Fig.~\ref{OS2}c), $M_2$ (Fig.~\ref{OS2}d),
and a single difficult integral $N(1,1,1,0,0)$ (Fig.~\ref{OS2}d),
with rational coefficients.
Instead of using $N(1,1,1,0,0)$ as a basis integral,
it is more convenient to use the convergent integral $N(1,1,1,1,1)$.
It can be expressed via ${}_3\!F_2$ hypergeometric functions
of the unit argument with indices depending on $\epsilon$.
Several terms of expansion in $\epsilon$ are known~\cite{B}:
\begin{equation}
N(1,1,1,1,1) = \pi^2 \log 2 - \tfrac{3}{2} \zeta(3) + \mathcal{O}(\epsilon)\,.
\label{N1}
\end{equation}

\begin{figure}[ht]
\begin{center}
\begin{picture}(52,20)
\put(26,10){\makebox(0,0){\includegraphics{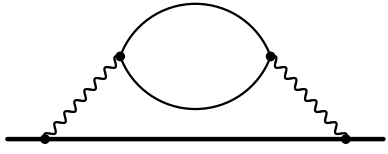}}}
\end{picture}
\end{center}
\caption{Two-loop on-shell propagator diagram with two masses}
\label{OSMm}
\end{figure}

Until now, we discussed on-shell propagator diagrams with a single non-zero mass.
Starting from two loops, there are also diagrams with loops of a different
massive quark, say, $c$ quark loops in $b$ quark self-energy, or vice versa
(Fig.~\ref{OSMm}).
Such diagrams can be reduced~\cite{DG}, using integration by parts,
to two trivial integrals (products of one-loop ones)
and two non-trivial integrals.
They are expressed via ${}_3\!F_2$ hypergeometric functions
of the mass ratio squared;
their finite terms at $\epsilon\to0$ are expressed via dilogarithms
of the mass ratio.

The first three-loop on-shell calculation,
that of the electron anomalous magnetic moment in QED,
has been completed recently~\cite{LR}.
A systematic algorithm for calculation of three-loop on-shell propagator diagrams
in QED and QCD, using integration by parts,
was constructed and implemented in~\cite{MR}.

\begin{figure}[ht]
\begin{center}
\begin{picture}(22,4)
\put(11,2){\makebox(0,0){\includegraphics{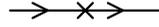}}}
\end{picture}
\end{center}
\caption{Mass counterterm vertex}
\label{CTVert}
\end{figure}

The on-shell renormalization scheme is most convenient for calculation
of on-shell scattering amplitudes.
The heavy-quark part of the QCD Lagrangian~(\ref{Lagr0}) can be rewritten as
\begin{equation}
L = \overline{Q}_0 (i\D_0-m) Q_0 + \delta m\, \overline{Q}_0 Q_0 + \cdots
\label{MassCT}
\end{equation}
where $m$ is the on-shell mass
(defined as the position of the pole of the full quark propagator),
and $\delta m=m-m_0$ is the mass counterterm
($m_0=Z_m^{\text{os}}m$).
We shall consider it not as a part of the unperturbed Lagrangian,
but as a perturbation.
It produces the counterterm vertex (Fig.~\ref{CTVert}) $i\,\delta m$.

The bare heavy-quark self energy can be decomposed as
\begin{equation}
\Sigma(p) = m \Sigma_1(p^2) + \left(\rlap/p-m\right) \Sigma_2(p^2)\,.
\label{SigOS}
\end{equation}
The bare heavy-quark propagator is then
\begin{equation}
S_0(p) = \frac{1}{\left(1-\Sigma_2(p^2)\right)\left(\rlap/p-m\right)+\delta m-m\Sigma_1(p^2)}\,.
\label{SOS}
\end{equation}
It has the pole at $p^2=m^2$ if
\begin{equation}
\delta m = m\Sigma_1(m^2)\,.
\label{OScond}
\end{equation}
The mass counterterm $\delta m$ is determined from this equation,
order by order in perturbation theory;
$Z_m^{\text{os}}=1-\delta m/m=1-\Sigma_1(m^2)$.
Then, near the mass shell, $\Sigma_1(p^2)-\delta m/m=\Sigma_1'(m^2)\,(p^2-m^2)+\cdots$,
where $\Sigma_1'(p^2)=d\Sigma_1(p^2)/dp^2$.
The bare quark propagator~(\ref{SOS}) becomes
\begin{equation}
S_0(p) = \frac{1}{1-\Sigma_2(m^2)-2m^2\Sigma_1'(m^2)} \frac{\rlap/p+m}{p^2-m^2} + \cdots
\label{Smshell}
\end{equation}
where dots mean terms which are non-singular at $p^2\to m^2$.
We define the heavy quark field renormalization constant in the on-shell scheme $Z_Q^{\text{os}}$
by the requirement that the renormalized quark propagator $S(p)$
(which is related to $S_0(p)$ by $S_0(p)=Z_Q^{\text{os}} S(p)$)
behaves as the free one~(\ref{Quark0}) near the mass shell.
Therefore, $Z_Q^{\text{os}}=\left[1-\Sigma_2(m^2)-2m^2\Sigma_1'(m^2)\right]^{-1}$.

Now let's explicitly calculate $Z_m^{\text{os}}$ and $Z_Q^{\text{os}}$
at the one-loop order.
It is convenient~\cite{MR} to introduce the function
\begin{equation}
T(t) = \frac{1}{4m} \Tr (\rlap/v+1) \Sigma(mv(1+t)) =
\Sigma_1(m^2) + \left[ \Sigma_2(m^2) + 2 m^2 \Sigma_1'(m^2) \right] t + \mathcal{O}(t^2)\,,
\label{Tt}
\end{equation}
then $Z_m^{\text{os}}=1-T(0)$ and $Z_Q^{\text{os}}=\left[1-T'(0)\right]^{-1}$.
We have
\begin{equation*}
T(t) = -i C_F g_0^2 \int \frac{d^d k}{(2\pi)^d} \frac{1}{D_1(t) D_2}
\frac{1}{4m} \Tr (\rlap/v+1) \gamma^\mu (\rlap/p+\rlap/k+m) \gamma^\nu
\left[ g_{\mu\nu} + (1-a_0) \frac{k_\mu k_\nu}{D_2} \right]\,,
\end{equation*}
where $D_1(t)=m^2-(p+k)^2$ and $D_2=-k^2$.
While calculating the numerator, we can express $p\cdot k$ via $D_1(t)$ and $D_2$,
and omit terms with $D_1(t)$, because resulting integrals are no-scale.
Omitting also $t^2$ and higher terms, we obtain
\begin{equation*}
T(t) = -i C_F g_0^2 \int \frac{d^d k}{(2\pi)^d} \frac{1}{D_1(t)}
\left[ \frac{2}{D_2} - \frac{d-2}{2m^2} (1-t) \right]\,.
\end{equation*}
Note that this result is gauge-independent.
Now, taking into account $D_1(t)=D_1+(D_1-D_2-2m^2)t+\mathcal{O}(t^2)$,
we arrive at
\begin{equation*}
T(t) =
C_F \frac{g_0^2 m^{-2\epsilon}}{(4\pi)^{d/2}} \Gamma(\epsilon) \frac{d-1}{d-3} (1-t)
+ \mathcal{O}(t^2)\,.
\end{equation*}
Therefore,
\begin{equation}
Z_m^{\text{os}} = Z_Q^{\text{os}} = 1 - C_F \frac{g_0^2 m^{-2\epsilon}}{(4\pi)^{d/2}}
\Gamma(\epsilon) \frac{d-1}{d-3}\,.
\label{Zos}
\end{equation}
The equality $Z_m^{\text{os}}=Z_Q^{\text{os}}$ is accidental,
and does not hold at higher orders.

On-shell renormalization of QCD at two loops has been performed in~\cite{B}
(see also~\cite{DG} for exact $d$-dimensional contributions of loops of another
massive quark).
The $\mathcal{O}(g_0^2)$ term in $\delta m=-m(Z_m^{\text{os}}-1)$, found in~(\ref{Zos}),
is necessary when calculating $\mathcal{O}(g_0^4)$ diagrams containing
the counterterm vertex (Fig.~\ref{CTVert}).
Three-loop results have been obtained recently~\cite{MR}.
The on-shell mass is gauge invariant to all orders~\cite{Kr};
the quark field renormalization $Z_Q^{\text{os}}$
is gauge invariant at two loops~\cite{B} but not at three~\cite{MR}.

\section{On-shell renormalization of HQET}
\label{SecHQETos}

\begin{figure}[ht]
\begin{center}
\begin{picture}(52,20)
\put(26,10){\makebox(0,0){\includegraphics{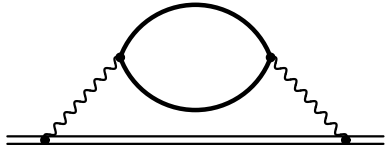}}}
\end{picture}
\end{center}
\caption{Two-loop on-shell HQET propagator diagram}
\label{OSHQET}
\end{figure}

Now we shall consider on-shell renormalization of the HQET Lagrangian~(\ref{LagrHQET0}).
HQET mass shell is $\omega=0$.
All loop diagrams without massive particles are no-scale and hence vanish.
Only diagrams with loops of massive quarks (of other flavours) can contribute.
Such diagrams first appear at two loops (Fig.~\ref{OSHQET}).

\begin{figure}[ht]
\begin{center}
\begin{picture}(22,22)
\put(11,11){\makebox(0,0){\includegraphics{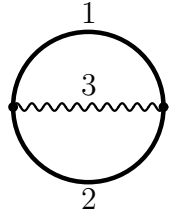}}}
\put(11,22){\makebox(0,0)[b]{1}}
\put(11,0){\makebox(0,0)[t]{2}}
\put(11,12.5){\makebox(0,0)[b]{3}}
\end{picture}
\end{center}
\caption{Two-loop vacuum diagram}
\label{Vac}
\end{figure}

The following method is used for calculation of such diagrams~\cite{BG3,CG}.
For two vectors $a$ and $b$ in $d$-dimensional Euclidean space,
the average over directions of $a$ (or $b$)
\begin{equation}
\overline{(a\cdot b)^n} = (a^2 b^2)^{n/2}
\frac{\int_0^\pi \cos^n\theta\,\sin^{d-2}\theta\,d\theta}
{\int_0^\pi \sin^{d-2}\theta\,d\theta} =
\frac{\Gamma\left(\frac{n+1}{2}\right)}{\Gamma\left(\frac{1}{2}\right)}
\frac{\Gamma\left(\frac{d}{2}\right)}{\Gamma\left(\frac{d+n}{2}\right)}
(a^2 b^2)^{n/2}
\label{av}
\end{equation}
for even $n$ (positive or negative), and 0 for odd $n$.
In particular, in a $d=1$ dimensional space the right-hand side is just $(a^2 b^2)^{n/2}$,
as expected.
We can use this formula for $\overline{(k\cdot v)^n}$ in Minkowski scalar integrals,
because they are calculated via Wick rotation.
The HQET propagator in Fig.~\ref{OSHQET} produces just an additional power of $k^2$,
and we are left with the vacuum diagram of Fig.~\ref{Vac}.
This diagram has been calculated in~\cite{Vl}:
\begin{equation}
\begin{split}
&\hspace{-5mm}\int \frac{d^d k_1\, d^d k_2}{(m^2-k_1^2-i0)^{n_1}(m^2-k_2^2-i0)^{n_2}
\left[-(k_1-k_2)^2-i0\right]^{n_3}} = - \pi^d m^{2(d-n_1-n_2-n_3)}\\
&\hspace{-5mm}\times
\frac{\Gamma(-d/2+n_1+n_3)\Gamma(-d/2+n_2+n_3)\Gamma(d/2-n_3)\Gamma(-d+n_1+n_2+n_3)}
{\Gamma(n_1)\Gamma(n_2)\Gamma(d/2)\Gamma(-d+n_1+n_2+2n_3)}\,.
\end{split}
\label{vac}
\end{equation}

Using this method, we see that $\Si(0)=0$, and
\begin{align}
\left.\frac{d\Si(\omega)}{d\omega}\right|_{\omega=0} &=
-i C_F g_0^2 \int \frac{d^d k}{(2\pi)^d} \frac{v^\mu v^\nu}{(k\cdot v)^2}
\frac{\Pi_{\mu\nu}(k)}{(k^2)^2} =
-i C_F g_0^2 \int \frac{d^d k}{(2\pi)^d} \left( \frac{k^2}{(k\cdot v)^2} - 1 \right)
\frac{\Pi(k^2)}{(k^2)^2}
\nonumber\\
&= - C_F T_F \frac{g_0^4 \sum m_i^{-4\epsilon}}{(4\pi)^d} \Gamma^2(\epsilon)
\frac{2(d-1)(d-6)}{(d-2)(d-5)(d-7)}\,,
\label{Sipr}
\end{align}
where $\Pi_{\mu\nu}(k)=(k^2 g_{\mu\nu}-k_\mu k_\nu)\Pi(k^2)$
is the massive quark contribution to the gluon self-energy,
$\overline{k^2/(k\cdot v)^2}=-(d-2)$,
and the sum is over all massive flavours.
We find the on-shell HQET quark field renormalization constant
$\Z_Q^{\text{os}}=\bigl[1-(d\Si(\omega)/d\omega)_{\omega=0}\bigr]^{-1}$
from the requirement that the renormalized propagator
$\S=\S_0/\Z_Q^{\text{os}}$ (see~(\ref{HQETmshell}))
behaves as the free one $\S^0$ near the mass shell:
\begin{equation}
\Z_Q^{\text{os}} =
1 - C_F T_F \frac{g_0^4 \sum m_i^{-4\epsilon}}{(4\pi)^d} \Gamma^2(\epsilon)
\frac{2(d-1)(d-6)}{(d-2)(d-5)(d-7)}\,.
\label{ZHQETOS}
\end{equation}
This renormalization constant is not smooth at $m_i\to0$.
This discontinuity comes from infrared gluon momenta, where HQET does not differ from QCD.
Therefore, the QCD on-shell quark field renormalization constant $Z_Q^{\text{os}}$
has the same non-smooth behaviour at $m_i\to0$~\cite{B,DG}.

\begin{figure}[ht]
\begin{center}
\begin{picture}(146,20)
\put(73,10){\makebox(0,0){\includegraphics{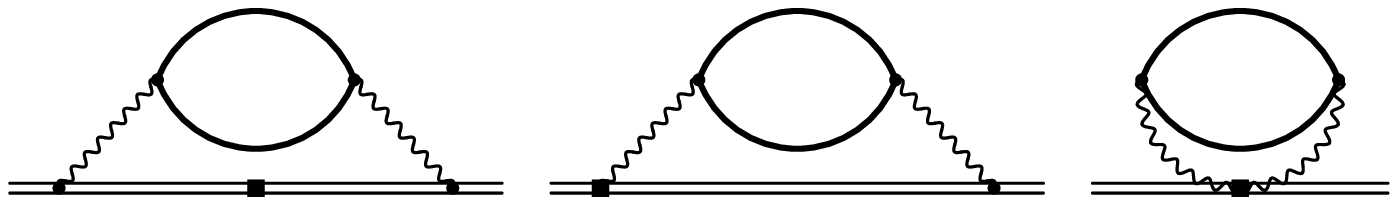}}}
\end{picture}
\end{center}
\caption{Two-loop diagrams for $\Si_{k0}(0)$}
\label{Sik0}
\end{figure}

Now we are going to calculate $\Si_{k0}(0)$ (see~(\ref{Repar3})).
At the two-loop level, it is given by the diagrams of Fig.~\ref{Sik0}
(where the second one also the mirror symmetric diagram),
with zero external residual momentum ($\omega=0$, $p_\bot=0$).
We obtain
\begin{equation}
\Si_{k0}(0) = i C_F g_0^2 \int \frac{d^d k}{(2\pi)^d} \frac{\Pi(k^2)}{k^2}
\left[ d-2 + \frac{k^2}{(k\cdot v)^2} \right] = 0\,.
\label{sik0}
\end{equation}

\section{Scattering in external gluonic field in QCD}
\label{SecExtQCD}

Now we are going to perform matching for the scattering amplitudes
of an on-shell heavy quark in QCD and in HQET,
with the linear accuracy in $q/m$,
where $q$ is the momentum transfer.

\begin{figure}[ht]
\begin{center}
\begin{picture}(98,28)
\put(49,14){\makebox(0,0){\includegraphics{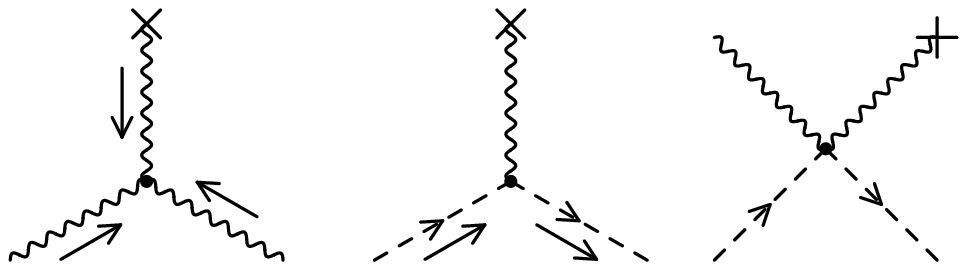}}}
\put(15,-5){\makebox(0,0)[b]{$a$}}
\put(12,29){\makebox(0,0)[t]{$\mu_1$}}
\put(18,29){\makebox(0,0)[t]{$a_1$}}
\put(0,3.5){\makebox(0,0){$a_2$}}
\put(3,-1){\makebox(0,0){$\mu_2$}}
\put(30,3.5){\makebox(0,0){$\mu_3$}}
\put(27,-1){\makebox(0,0){$a_3$}}
\put(9.5,17){\makebox(0,0){$k_1$}}
\put(10.5,-0.5){\makebox(0,0){$k_2$}}
\put(24.5,9){\makebox(0,0){$k_3$}}
\put(52,-5){\makebox(0,0)[b]{$b$}}
\put(37,3.5){\makebox(0,0){$b_1$}}
\put(67,3.5){\makebox(0,0){$b_2$}}
\put(49,29){\makebox(0,0)[t]{$\mu$}}
\put(55,29){\makebox(0,0)[t]{$a$}}
\put(47.5,-0.5){\makebox(0,0){$k_1$}}
\put(56.5,-0.5){\makebox(0,0){$k_2$}}
\put(84,-5){\makebox(0,0)[b]{$c$}}
\put(71,3.5){\makebox(0,0){$b_1$}}
\put(97,3.5){\makebox(0,0){$b_2$}}
\put(70.5,21.5){\makebox(0,0){$a_1$}}
\put(97.5,21.5){\makebox(0,0){$a_2$}}
\put(75,25.5){\makebox(0,0){$\mu_1$}}
\put(93,25.5){\makebox(0,0){$\mu_2$}}
\end{picture}
\end{center}
\caption{Vertices of interaction with the background field}
\label{BFvert}
\end{figure}

It is most convenient to calculate scattering amplitudes in an external field
using the background field method~\cite{Ab}.
In the QCD Lagrangian~(\ref{Lagr0}),
we substitute $A_0^\mu\to\overline{A}_0^\mu+A_0^\mu$,
where $\overline{A}_0^\mu$ is the external field,
and choose the gauge fixing term $\left(\overline{D}_\mu A_0^\mu\right)^2/(2a_0)$,
where $\overline{D}_\mu=\partial_\mu-ig_0\overline{A}_{0\mu}^a t^a$.
The ghost term is changed correspondingly:
\begin{equation}
L = \sum_i \overline{q}_{i0} (i\D_0-m_{i0}) q_{i0}
- \frac{1}{4} G^a_{0\mu\nu} G^{a\mu\nu}_0
- \frac{1}{2a_0} \left(\overline{D}_\mu A_0^\mu\right)^2
+ (\overline{D}_\mu \overline{c}_0^a) (D_0^\mu c_0^a)\,.
\label{LagrBF}
\end{equation}
Some vertices with the background field $\overline{A}_0$
differ from the ordinary ones.
In particular, the gauge fixing term contains a $\overline{A}_0 A_0^2$ contribution,
altering the three-gluon vertex (Fig.~\ref{BFvert}a) to
\begin{equation*}
g_0 f^{a_1 a_2 a_3} \left[ \left(k_2-k_3\right)^{\mu_1} g^{\mu_2\mu_3}
+ \left(k_3-k_1+\frac{1}{a_0}k_2\right)^{\mu_2} g^{\mu_3\mu_1}
+ \left(k_1-k_2-\frac{1}{a_0}k_3\right)^{\mu_3} g^{\mu_1\mu_2} \right]\,.
\end{equation*}
It contains no $\overline{A}_0 A_0^3$ contribution,
so that the four-gluon vertex does not change.
The ghost term in~(\ref{LagrBF}) gives the vertices (Fig.~\ref{BFvert}b, c)
\begin{equation*}
-g_0 f^{a b_1 b_2} (k_1+k_2)^\mu\,, \quad
i g_0^2 f^{a_1 b_1 c} f^{a_2 b_2 c} g^{\mu_1\mu_2}\,.
\end{equation*}
The terms with $1/a_0$ in the three-gluon vertex contain $k_1^{\mu_1}$ or $k_2^{\mu_2}$;
when they are multiplied by the propagator $D_{0\mu_1\nu}(k_1)$ or $D_{0\mu_2\nu}(k_2)$,
correspondingly, they extract the term with $a_0$ from the propagator~(\ref{Glue0}),
and no terms with negative powers of the gauge parameter $a_0$ appear.

\begin{figure}[p]
\begin{center}
\begin{picture}(60,24)
\put(30,12){\makebox(0,0){\includegraphics{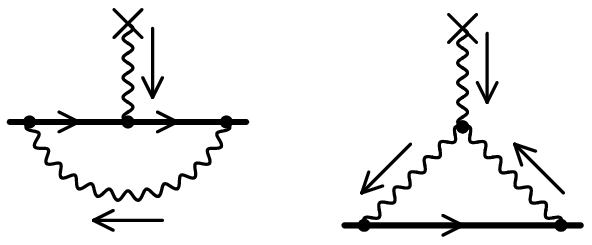}}}
\put(13,-6){\makebox(0,0)[b]{$a$}}
\put(13,-1){\makebox(0,0){$k$}}
\put(7,15){\makebox(0,0){$k+p$}}
\put(24,15){\makebox(0,0){$k+p+q$}}
\put(17.5,20.5){\makebox(0,0){$q$}}
\put(47,-6){\makebox(0,0)[b]{$b$}}
\put(47,-1){\makebox(0,0){$k+p$}}
\put(37,10){\makebox(0,0){$k$}}
\put(60,10){\makebox(0,0){$k-q$}}
\put(51.5,20.5){\makebox(0,0){$q$}}
\end{picture}
\end{center}
\caption{One-loop proper vertex}
\label{Vert1Loop}
\end{figure}

The sum of one-particle-irreducible vertex diagrams
not including the external propagators
is the proper vertex $i g_0 t^a \Gamma^\mu(p,q)$,
where $p$ is the incoming quark momentum,
$p'=p+q$ is the outgoing quark momentum,
$\mu$ and $a$ are the background-field gluon polarization and colour indices.
The vertex function is $\Gamma^\mu(p,q)=\gamma^\mu+\Lambda^\mu(p,q)$,
where $\Lambda^\mu(p,q)$ contains one-loop (Fig.~\ref{Vert1Loop})
and higher-loop corrections.

\begin{figure}[p]
\begin{center}
\begin{picture}(107,32)
\put(53.5,16){\makebox(0,0){\includegraphics{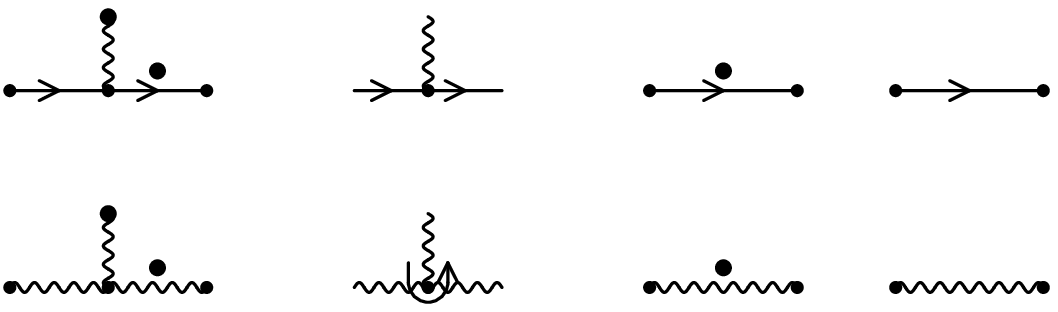}}}
\put(53.5,14){\makebox(0,0)[b]{$a$}}
\put(28.5,23){\makebox(0,0){${}=g_0$}}
\put(58.5,23){\makebox(0,0){${}\times\Biggl[\Biggr.$}}
\put(86,23){\makebox(0,0){$-$}}
\put(111,23){\makebox(0,0){$\Biggl.\Biggr]$}}
\put(53.5,-6){\makebox(0,0)[b]{$b$}}
\put(28.5,3){\makebox(0,0){${}=g_0$}}
\put(58.5,3){\makebox(0,0){${}\times\Biggl[\Biggr.$}}
\put(86,3){\makebox(0,0){$-$}}
\put(111,3){\makebox(0,0){$\Biggl.\Biggr]$}}
\end{picture}
\end{center}
\caption{Ward identities for the background-field vertices}
\label{WardBF}
\end{figure}

\begin{figure}[p]
\begin{center}
\begin{picture}(128,77)
\put(64,38.5){\makebox(0,0){\includegraphics{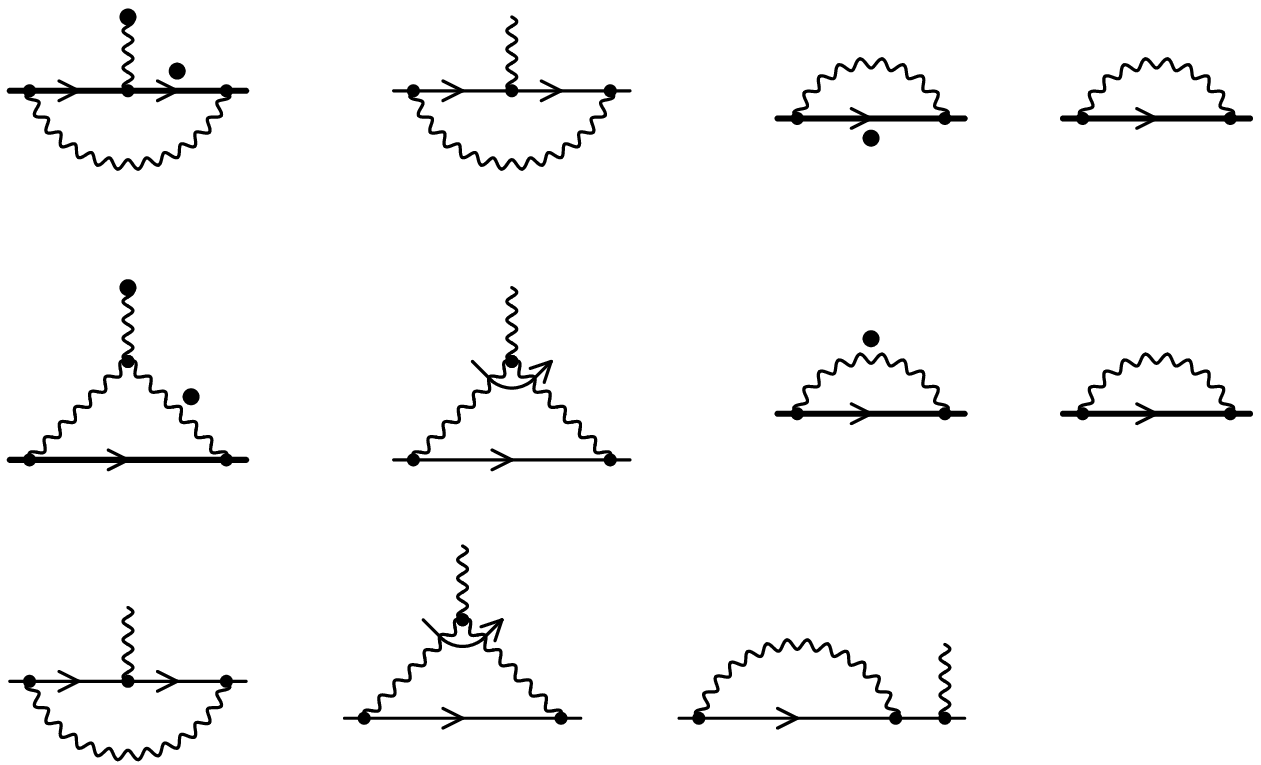}}}
\put(64,55){\makebox(0,0)[b]{$a$}}
\put(32.5,68.5){\makebox(0,0){${}=g_0$}}
\put(71.5,68.5){\makebox(0,0){$\times\Biggl[\Biggr.$}}
\put(103,68.5){\makebox(0,0){$-$}}
\put(132,68.5){\makebox(0,0){$\Biggl.\Biggr]$}}
\put(64,25){\makebox(0,0)[b]{$b$}}
\put(32.5,38.5){\makebox(0,0){${}=g_0$}}
\put(71.5,38.5){\makebox(0,0){$\times\Biggl[\Biggr.$}}
\put(103,38.5){\makebox(0,0){$-$}}
\put(132,38.5){\makebox(0,0){$\Biggl.\Biggr]$}}
\put(64,-5){\makebox(0,0)[b]{$c$}}
\put(30,8.5){\makebox(0,0){$+$}}
\put(64,8.5){\makebox(0,0){$=$}}
\end{picture}
\end{center}
\caption{Ward identity for the background-field vertex function}
\label{WardBFV}
\end{figure}

Now we are going to derive the Ward identity for $\Lambda^\mu(p,q)q_\mu$.
Background-field vertices obey simple identities shown in Fig.~\ref{WardBF},
where the colour structure is singled out as the first factor,
the fat dot at the end of the background gluon line means multiplication by $q_\mu$,
a dot near a propagator means that its momentum is shifted by $q$
(as in Fig.~\ref{WardS}).
Starting from each diagram for $\Sigma(p)$,
we can obtain a set of diagrams for $\Lambda^\mu$
by attaching the background-field gluon to each possible place.
For example, starting from the one-loop diagram Fig.~\ref{Quark1Loop} for $\Sigma(p)$,
we can attach the external gluon either to the quark line, or to the gluon one,
obtaining the diagrams Fig.~\ref{Vert1Loop}.
Results of their multiplication by $q_\mu$ are shown in Fig.~\ref{WardBFV}a, b.
The differences in the square brackets are equal.
The colour factors combine to give the colour factor of Fig.~\ref{Quark1Loop}
times $t^a$ (Fig.~\ref{WardBFV}c), due to the definition of the colour factor
of the three-gluon vertex (Fig.~\ref{Cvit4}).
Therefore, we have the Ward identity
\begin{equation}
\Lambda^\mu(p,q) q_\nu = \Sigma(p) - \Sigma(p+q) \quad \text{or} \quad
\Gamma^\mu(p,q) q_\mu = S_0^{-1}(p+q) - S_0^{-1}(p)\,.
\label{wardBF}
\end{equation}
It also holds at higher orders of perturbation theory.
To verify this, one has to derive identities similar to Fig.~\ref{WardBF}
for other background-field vertices.
For an infinitesimal $q$, we have
\begin{equation}
\Lambda^\mu(p,0) = - \frac{\partial \Sigma(p)}{\partial p_\mu} \quad \text{or} \quad
\Gamma^\mu(p,0) = \frac{\partial S_0^{-1}(p)}{\partial p_\mu}\,.
\label{ward0}
\end{equation}

The Ward identities~(\ref{wardBF}), (\ref{ward0}) are very simple,
exactly the same as in QED
(or in the Heavy Electron Effective Theory, Sect.~\ref{SecHEET}).
Multiplying the ordinary three-gluon vertex by $q_\mu$ gives an identity
which, in addition to the simple difference of Fig.~\ref{WardBF}b,
has additional ghost terms.
Therefore, the Ward identities for the ordinary quark-gluon vertex function
are more complicated than~(\ref{wardBF}), (\ref{ward0}).

The renormalized background field is related to the bare one as
$\overline{A}_0=\overline{Z}_A^{1/2}\overline{A}$.
The renormalized matrix element is the proper vertex $g_0 t^a \Gamma^\mu(p,q)$
times $Z_q \overline{Z}_A^{1/2}$:
$Z_q \overline{Z}_A^{1/2} Z_\alpha^{1/2} g t^a \Gamma^\mu(p,q)$.
At $q=0$, the factor $Z_q$ converts $S_0^{-1}$ in~(\ref{ward0})
into $S^{-1}$, making it finite.
Therefore, $Z_\alpha \overline{Z}_A=1$, just as in QED
(or in Heavy Electron Effective Theory, Sect.~\ref{SecHEET}).
In other words, $g_0\overline{A}_0=g\overline{A}$.

Scattering amplitude of the on-shell quark in an external field is
$g Z_Q^{\text{os}} \overline{u}(p')\Gamma^\mu t^a u(p)$,
where $p^2=p^{\prime2}=m^2$, $(\rlap/p-m)u(p)=0$, $(\rlap/p'-m)u(p')=0$.
It is UV finite, but may contain IR divergences.
It can be expressed via two scalar form factors.
For comparing with HQET, it is most convenient to use
the Dirac and chromomagnetic form factors:
\begin{equation}
\overline{u}(p') \Gamma^\mu t^a u(p) =
\overline{u}(p') \left( F_1^0(q^2) \frac{(p+p')^\mu}{2m}
+ G_m^0(q^2) \frac{[\rlap/q,\gamma^\mu]}{4m} \right) t^a u(p)\,,
\label{FF}
\end{equation}
where the renormalized form factors are $F_1(q^2)=Z_Q^{\text{os}}F_1^0(q^2)$,
$G_m(q^2)=Z_Q^{\text{os}}G_m^0(q^2)$.
The form factors can be singled out by the appropriate projectors.
Let's rewrite~(\ref{FF}) as
\begin{align*}
&\overline{u}(p') \Gamma^\mu t^a u(p) = \overline{u}(p') \sum F_i T_i^\mu\, t^a u(p)\,,\\
&F_1 = F_1^0(q^2)\,, \quad
F_2 = G_m^0(q^2)\,, \quad
T_1^\mu = \frac{(p+p')^\mu}{2m}\,, \quad
T_2^\mu = \frac{[\rlap/q,\gamma^\mu]}{4m}\,,
\end{align*}
and calculate the traces
\begin{equation*}
\frac{1}{4} \Tr \Gamma_\mu (\rlap/v+1) T_i^\mu \left(\rlap/v+\rlap/q/m+1\right)\,.
\end{equation*}
Solving the linear system for $F_i$, we obtain
\begin{align}
F_1^0(q^2) =& \frac{1}{2(d-2)\left(1-\frac{q^2}{4m^2}\right)}
\nonumber\\
&{}\times \frac{1}{4} \Tr \Gamma_\mu (\rlap/v+1)
\left( \frac{d-2+\frac{q^2}{4m^2}}{1-\frac{q^2}{4m^2}} \left(v+\frac{q}{2m}\right)^\mu
+ \frac{[\rlap/q,\gamma^\mu]}{4m} \right) \left(\rlap/v+\frac{\rlap/q}{m}+1\right)\,,
\label{Pr}\\
G_m^0(q^2) =& - \frac{1}{2(d-2)} \frac{1}{4} \Tr \Gamma_\mu (\rlap/v+1)
\left( \frac{1}{1-\frac{q^2}{4m^2}} \left(v+\frac{q}{2m}\right)^\mu
+ \frac{[\rlap/q,\gamma^\mu]}{q^2} \right) \left(\rlap/v+\frac{\rlap/q}{m}+1\right)\,.
\nonumber
\end{align}
There is no singularity in $G_m^0(q^2)$ at $q^2\to0$.
We can expand the form factors in $q^2$ expanding
\begin{equation*}
\Gamma^\mu = \Gamma_0^\mu + \Gamma_1^{\mu\nu} \frac{q_\nu}{m} + \cdots\,,
\end{equation*}
splitting $q=(q\cdot v)v+q_\bot$ (where $q\cdot v=-q^2/(2m)$)
and averaging over directions of $q_\bot$
in the $(d-1)$-dimensional subspace orthogonal to $v$:
\begin{equation*}
\overline{q^\alpha} = - \frac{q^2}{2m} v^\alpha\,,\quad
\overline{q^\alpha q^\beta} = \frac{q^2}{d-1}
\left[ \left(1-\frac{q^2}{4m^2}\right) g^{\alpha\beta}
- \left(1-d\frac{q^2}{4m^2}\right) v^\alpha v^\beta \right]\,,
\quad\ldots
\end{equation*}
Thus we obtain
\begin{align}
F_1^0(q^2) =& \frac{1}{4} \Tr \Gamma_0^\mu (\rlap/v+1) v_\mu + \mathcal{O}(q^2/m^2)\,,
\nonumber\\
G_m^0(q^2) =& \frac{1}{d-1} \frac{1}{4} \Tr \Gamma_0^\mu (\rlap/v+1) (\gamma_\mu-v_\mu)
\label{Pr0}\\
&{} + \frac{2}{(d-1)(d-2)} \frac{1}{4} \Tr \Gamma_1^{\mu\nu} (\rlap/v+1)
(\gamma_\mu\gamma_\nu - g_{\mu\nu} + \gamma_\mu v_\nu - \gamma_\nu v_\mu)
+ \mathcal{O}(q^2/m^2)\,.
\nonumber
\end{align}

The Dirac form factor at $q=0$ is unity, due to the Ward identity~(\ref{ward0}):
\begin{align}
F_1(0) &= Z_Q^{\text{os}} \left[ 1 + \frac{1}{4} \Tr \Lambda_\mu(mv,0) (\rlap/v+1) v^\mu \right]
= Z_Q^{\text{os}} \left[ 1 - v^\mu \frac{\partial}{\partial p^\mu} \left.
\frac{1}{4} \Tr \Sigma(p) (\rlap/v+1) \right|_{p=mv} \right]
\nonumber\\
&= Z_Q^{\text{os}} \left[ 1 - T'(0) \right] = 1
\label{F10}
\end{align}
(see~(\ref{Tt})).
The total colour charge of the heavy quark is not changed by renormalization.

The heavy-quark chromomagnetic moment $\mu_g=G_m(0)$~(\ref{Pr0})
can be calculated as follows.
First, we apply the projector in~(\ref{Pr0}) to the integrand of each diagram,
differentiating all $q$-dependent propagators and vertices (Fig.~\ref{Vert1Loop}),
and obtain the bare $\mu_g^0$ via scalar integrals~(\ref{M1}).
Then we multiply it by $Z_Q^{\text{os}}$~(\ref{Zos}).
The one-loop result is~\cite{EH2}:
\begin{equation}
\mu_g = 1 + \frac{g_0^2 m^{-2\epsilon}}{(4\pi)^{d/2}} \frac{\Gamma(\epsilon)}{2(d-3)}
\left[ 2(d-4)(d-5) C_F - (d^2-8d+14) C_A \right] + \cdots
\label{mug}
\end{equation}
Setting $g_0\to e_0$, $C_F\to1$, $C_A\to0$,
we reproduce the electron magnetic moment $\mu=1+\frac{\alpha}{2\pi}+\cdots$ in QED.
It is convergent.
The heavy-quark chromomagnetic moment in QCD~(\ref{mug})
contains an IR divergence with the colour factor $C_A$.
The two-loop correction to~(\ref{mug}) was calculated in~\cite{CG},
and the effect of another massive flavour
(say, $c$ loop correction to the $b$ quark chromomagnetic moment)
was considered in~\cite{DG}.

\section{Scattering in external gluonic field in HQET}
\label{SecExtHQET}

First, let's consider the leading (zeroth) order in $1/m$.
Let the sum of one-particle-irreducible bare vertex diagrams
in the background-field method be $i g_0 t^a \Ga^\mu(\omega,q)$,
where $\omega$ and $\omega'=\omega+q\cdot v$ are the residual energies
of the initial and the final quark;
$\Ga^\mu(\omega,q)=v^\mu+\La^\mu(\omega,q)$.
There are two vectors in the problem, $v^\mu$ and $q^\mu$,
and hence the vertex function has the structure
\begin{equation}
\begin{split}
&\La^\mu = \La_s(\omega,\omega',q^2) v^\mu
+ \La_a(\omega,\omega',q^2) q^\mu\,,\\
&\La_s(\omega,\omega',q^2) = \La_s(\omega',\omega,q^2)\,,\quad
\La_a(\omega,\omega',q^2) = -\La_a(\omega',\omega,q^2)\,.
\end{split}
\label{Las}
\end{equation}
It obeys the Ward identity
\begin{equation}
\La^\mu(\omega,q) q_\nu = \La_s(\omega'-\omega) + \La_a q^2 = \Si(\omega) - \Si(\omega')
\quad \text{or} \quad
\Ga^\mu(\omega,q) q_\mu = \S_0^{-1}(\omega') - \S_0^{-1}(\omega)\,,
\label{wardBFHQET}
\end{equation}
or, for $q\to0$,
\begin{equation}
\La_s(\omega,\omega,0) = - \frac{d\Si(\omega)}{d\omega}\,.
\label{ward0HQET}
\end{equation}
On the mass shell $\omega=0$, $\omega'=0$,
the renormalized scattering amplitude with the linear accuracy in $q$ is
\begin{equation}
\Z_Q^{\text{os}} (1+\Lambda_s(0,0,0)) \overline{u}_v(q) v^\mu u_v(0)
= \Z_Q^{\text{os}} \left(1 - \frac{d\Si(\omega)}{d\omega}\right)_{\omega=0}
\overline{u}_v(q) v^\mu u_v(0) = \overline{u}_v(q) v^\mu u_v(0)\,.
\label{SHQET0}
\end{equation}
The total colour charge of the static quark is not changed by renormalization.

There are two kinds of $1/m$ corrections to the scattering amplitude:
diagrams with a single kinetic vertex, and those with a single chromomagnetic vertex.
Let's forget for a while that we have obtained the result $C_k=1$~(\ref{ReparK}),
and denote the sum of one-particle-irreducible bare vertex diagrams
in the background-field method containing a single kinetic-energy vertex
$i g_0 t^a \frac{C_k^0}{2m} \Ga_k^\mu(\p,q)$,
where $\p=\omega v+p_\bot$ and $\p'=\p+q=\omega'v+p'_\bot$
are the initial and final residual momenta of the heavy quark;
$\Ga_k=(p+p')_\bot^\mu+\La_k^\mu(\p,q)$.
Dependence on $p_\bot$ and $p'_\bot$ comes only from the kinetic-energy vertex
(Sect.~\ref{Sec1m}), which is at most quadratic in them.
Therefore, the vertex function has the structure
\begin{align}
&\begin{split}
\La_k^\mu(\p,q) =& \La_{ks}(\omega,\omega',q^2) (p+p')_\bot^\mu\\
&{} + \left[
\La_{k1}(\omega,\omega',q^2) p_\bot^2 + \La_{k1}(\omega',\omega,q^2) p_\bot^{\prime2}
+ \La_{k0}(\omega,\omega',q^2) \right] v^\mu\\
&{} + \left[
\La_{k3}(\omega,\omega',q^2) p_\bot^2 - \La_{k3}(\omega',\omega,q^2) p_\bot^{\prime2}
+ \La_{k2}(\omega,\omega',q^2) \right] q^\mu\,,\\
\end{split}
\label{Lask}\\
&\begin{split}
&\La_{ks}(\omega,\omega',q^2) = \La_{ks}(\omega',\omega,q^2)\,,\\
&\La_{k0}(\omega,\omega',q^2) = \La_{k0}(\omega',\omega,q^2)\,,\quad
\La_{k2}(\omega,\omega',q^2) = -\La_{k2}(\omega',\omega,q^2)\,.
\end{split}
\nonumber
\end{align}

Similarly to Sec.~\ref{Sec1m}, the variation of diagrams for $\La(\p,q)$ at $v\to v+\delta v$
is equal to that of diagrams for $\La_k(\p,q)$ at $p_\bot\to p_\bot+\delta p_\bot$,
if $\delta v=\frac{C_k}{m}\delta p_\bot$.
Therefore, the reparametrization invariance ensures
\begin{align}
&\La_{ks}(\omega,\omega',q^2) = \La_s(\omega,\omega',q^2)\,,
\nonumber\\
&\La_{k1}(\omega,\omega',q^2) = \frac{\partial\La_s(\omega,\omega',q^2)}{\partial\omega}\,,
\label{ReparVk}\\
&\La_{k3}(\omega,\omega',q^2) = \frac{\partial\La_a(\omega,\omega',q^2)}{\partial\omega}\,.
\nonumber
\end{align}
The Ward identity
\begin{equation}
\La_k^\mu(\p,q) q_\mu = \Si_k(\p) - \Si_k(\p')
\label{wardVk}
\end{equation}
ensures, due to~(\ref{Repar3}),
\begin{align}
&{} - \La_{ks}(\omega,\omega',q^2) + \La_{k1}(\omega,\omega',q^2) (\omega'-\omega)
+ \La_{k3}(\omega,\omega',q^2) q^2 = \frac{d\Si(\omega)}{d\omega}\,,
\nonumber\\
&\La_{k0}(\omega,\omega',q^2) (\omega'-\omega) + \La_{k2}(\omega,\omega',q^2) q^2
= \Si_{k0}(\omega)- \Si_{k0}(\omega')\,.
\label{wardVk2}
\end{align}
The first relation here is, due to~(\ref{ReparVk}),
just the derivative of~(\ref{wardBFHQET}) in $\omega$.

On the mass shell $\omega=0$, $\omega'=0$, the kinetic-energy correction to
the renormalized scattering amplitude with the linear accuracy in $q$ is
\begin{equation}
\frac{C_k^0}{2m} \Z_Q^{\text{os}} (1+\Lambda_s(0,0,0)) \overline{u}_v(q) (p+p')_\bot^\mu u_v(0)
= \frac{C_k^0}{2m} \overline{u}_v(q) (p+p')_\bot^\mu u_v(0)\,.
\label{SHQETk}
\end{equation}

Let the sum of one-particle-irreducible bare vertex diagrams
in the background-field method containing a single chromomagnetic vertex be
\begin{equation}
i g_0 t^a \frac{C_m^0}{4m} \frac{1+\rlap/v}{2} [\rlap/q,\gamma^\mu] \frac{1+\rlap/v}{2}
\Ga_m(\omega,\omega',q^2)\,,
\label{VertCm}
\end{equation}
$\Ga_m(\omega,\omega',q^2)=1+\La_m(\omega,\omega',q^2)$.
Reparametrization invariance does not relate $\La_m$ with any vertex function
of zeroth order in $1/m$.
We have no better alternative to a direct calculation.
In order to obtain the on-shell scattering amplitude with the linear accuracy in $q$,
we only need the static quark chromomagnetic moment $\widetilde{\mu}_g^0=\Ga_m(0,0,0)$.
All loop diagrams for $\La_m$ vanish, except those with loops of some massive quark.
Such diagrams first appear at two loops.
They can be calculated using the method of Sect.~\ref{SecHQETos}.
The renormalized matrix element $\widetilde{\mu}_g=\Z_Q^{\text{os}}\widetilde{\mu}_g^0$
of the chromomagnetic operator $\widetilde{O}_m^0$ is~\cite{CG}
\begin{equation}
\widetilde{\mu}_g = 1 + C_A T_F \frac{g_0^4 \sum m_i^{-4\epsilon}}{(4\pi)^d}
\Gamma^2(\epsilon) \frac{d^2-9d+16}{(d-2)(d-5)(d-7)}\,,
\label{HQETmug}
\end{equation}
where the sum is over all massive flavours
(except the heavy flavour of our HQET, of course).

\section{Chromomagnetic interaction}
\label{SecCMag}

Now we are ready to compare the on-shell scattering amplitudes in full QCD and in HQET.
HQET spinors $u_v(\p)$ are 2-component in the $v$ rest frame: $\rlap/vu_v(\p)=0$.
The corresponding QCD spinors $u(mv+\p)$ are related to them by the Foldy-Wouthuysen
transformation~\cite{BD}
\begin{equation}
u(mv+\p) = \left(1 + \frac{\rlap/\p}{2m} + \mathcal{O}(\p^2/m^2) \right) u_v(\p)\,.
\label{FW}
\end{equation}
Expanding the QCD scattering amplitude up to linear terms in $q/m$
and re-expressing it via HQET spinors, we obtain
\begin{equation}
\overline{u}_v(q) \left( v^\mu + \frac{q^\mu}{2m}
+ \mu_g \frac{[\rlap/q,\gamma^\mu]}{4m} \right) t^a u_v(0)\,.
\label{SQCD}
\end{equation}
The HQET scattering amplitude with the $1/m$ accuracy is
\begin{equation}
\overline{u}_v(q) \left( v^\mu + C_k(\mu) \Z_k^{-1}(\mu) \frac{q^\mu}{2m}
+ C_m(\mu) \Z_m^{-1}(\mu) \widetilde{\mu}_g \frac{[\rlap/q,\gamma^\mu]}{4m}
\right) t^a u_v(0)\,.
\label{SHQET}
\end{equation}
Both scattering amplitudes~(\ref{SQCD}) and~(\ref{SHQET}) are renormalized
and hence UV finite.
Both may contain IR divergences.
By construction, HQET is identical to QCD in the infrared region,
so that these IR divergences are the same.
For example, if there are no other massive flavours in the theory,
all loop corrections to $\widetilde{\mu}_g^0$ vanish because they contain no scale.
These zero integrals contain UV and IR divergences which cancel.
UV divergences are removed by $\Z_m^{-1}(\mu)$, and IR ones match those in $\mu_g$.

Comparing the coefficients of $q^\mu/m$, we again see that $\Z_k=1$
and $C_k(\mu)=1$~(\ref{ReparK}).
Comparing the coefficients of $[\rlap/q,\gamma^\mu]/m$, we obtain
\begin{equation}
\Z_m^{-1}(\mu) C_m(\mu) = \frac{\mu_g}{\widetilde{\mu}_g}\,.
\label{MatchM}
\end{equation}
In the one-loop approximation, re-expressing~(\ref{mug}) via $\alpha_s(\mu)$~(\ref{alphas})
and expanding in $\epsilon$, we obtain
\begin{equation}
\Z_m^{-1}(\mu)C_m(\mu) = 1 + \frac{\alpha_s(\mu)}{4\pi} e^{-2L\epsilon}
\left[ 2 C_F + \left(\frac{1}{\epsilon}+2\right) C_A \right]\,, \quad
L = \log\frac{m}{\mu}\,.
\label{MatchM1}
\end{equation}
The minimal~(\ref{minim}) renormalization constant $\Z_m$ making $C_m$ finite is
\begin{equation}
\Z_m = 1 - C_A \frac{\alpha_s}{4\pi\epsilon} + \cdots\,,
\label{Zm1}
\end{equation}
and the chromomagnetic interaction constant is
\begin{equation}
C_m(\mu) = 1 + 2 \left( - C_A L + C_F + C_A \right) \frac{\alpha_s(m)}{4\pi} + \cdots
\label{Cm1}
\end{equation}
Therefore, the anomalous dimension of the chromomagnetic operator and $C_m(m)$ are~\cite{EH2}
\begin{equation}
\ga_m = 2 C_A \frac{\alpha_s}{4\pi} + \cdots\,, \quad
C_m(m) = 1 + 2(C_F + C_A) \frac{\alpha_s(m)}{4\pi} + \cdots
\label{gm1}
\end{equation}

Two-loop anomalous dimension was calculated in~\cite{ABN} within HQET,
and (a week later) in~\cite{CG} by QCD/HQET matching:
\begin{equation}
\ga_m = 2 C_A \frac{\alpha_s}{4\pi}
+ \frac{4}{9} C_A (17 C_A - 13 T_F n_f) \left(\frac{\alpha_s}{4\pi}\right)^2 + \cdots
\label{gm2}
\end{equation}
The anomalous dimension vanishes in QED, where $C_A=0$.
Two-loop correction to $C_m(m)$ was found in~\cite{CG}.
It is most convenient to calculate
\begin{equation*}
C_m(m)=1 + C_1 \frac{\alpha_s(m)}{4\pi} + C_2 \left(\frac{\alpha_s(m)}{4\pi}\right)^2 + \cdots
\end{equation*}
because it contains no large logs.
Then we can use the renormalization group equation to find $C_m(\mu)$ at $\mu\ll m$.

The scattering amplitude~(\ref{SHQET}) does not depend on an arbitrary
renormalization scale $\mu$;
the matrix element $\widetilde{\mu}_g^0$ of the bare chromomagnetic operator
is also $\mu$-independent.
Therefore,
\begin{equation*}
\Z_m^{-1}(\mu)C_m(\mu) = \text{const}\,.
\end{equation*}
Differentiating this equality in $\log\mu$, we obtain the renormalization group equation
\begin{equation}
\frac{d\,C_m(\mu)}{d\log\mu} = \ga_m(\alpha_s(\mu)) C_m(\mu)\,.
\label{RGm}
\end{equation}

If $L=\log m/\mu$ is not very large, it is reasonable to find the solution
as a series in $\alpha_s(m)$.
Re-expressing
\begin{equation*}
\frac{\alpha_s(\mu)}{4\pi} = \frac{\alpha_s(m)}{4\pi}
\left[ 1 + 2 \beta_0 L \frac{\alpha_s(m)}{4\pi} + \cdots \right]
\end{equation*}
in
\begin{equation*}
\ga_m(\alpha_s(\mu)) = \gamma_0 \frac{\alpha_s(\mu)}{4\pi}
+ \gamma_1 \left(\frac{\alpha_s(\mu)}{4\pi}\right)^2 + \cdots\,,
\end{equation*}
we obtain the equation
\begin{equation*}
\frac{d\,C_m(\mu)}{dL} + \left[ \gamma_0 \frac{\alpha_s(m)}{4\pi}
\left(1 + 2 \beta_0 L \frac{\alpha_s(m)}{4\pi}\right)
+ \gamma_1 \left(\frac{\alpha_s(m)}{4\pi}\right)^2 + \cdots \right] C_m(\mu)\,,
\end{equation*}
which can be solved order by order in $\alpha_s(m)$:
\begin{equation}
C_m(\mu) = 1 + (-\gamma_0 L + C_1) \frac{\alpha_s(m)}{4\pi}
+ \left[ \gamma_0 \left( \frac{\gamma_0}{2}-\beta_0\right) L^2
- \left(\gamma_1+C_1\gamma_0\right) L + C_2 \right]
\left(\frac{\alpha_s(m)}{4\pi}\right)^2 + \cdots
\label{sol1}
\end{equation}

If $\frac{\alpha_s}{4\pi}L\sim1$, we have to solve
the renormalization group equation~(\ref{RGm}) in another way.
Dividing it by $d\log\alpha_s(\mu)/d\log\mu$~(\ref{rengroup})
(at $\epsilon=0$), we obtain
\begin{equation*}
\frac{d\log C_m(\mu)}{d\log\alpha_s} + \frac{\ga_m(\alpha_s)}{2\beta(\alpha_s)} = 0\,.
\end{equation*}
The solution of this equation is
\begin{equation}
C_m(\mu) = C_m(m) \exp \left[ - \int\limits_{\alpha_s(m)}^{\alpha_s(\mu)}
\frac{\ga_m(\alpha_s)}{2\beta(\alpha_s)} \frac{d\alpha_s}{\alpha_s} \right] \,.
\label{sol}
\end{equation}
We can expand the ratio $\ga_m(\alpha_s)/\beta(\alpha_s)$ in $\alpha_s$.
Integral of the first term gives $\log\alpha_s(\mu)/\alpha_s(m)$.
Integrals of all the other terms are powers of $\alpha_s$,
and we may expand the exponent of these terms:
\begin{equation}
C_m(\mu) = \left(\frac{\alpha_s(\mu)}{\alpha_s(m)}\right)^{-\frac{\gamma_0}{2\beta_0}}
\left[1 + C_1 \frac{\alpha_s(m)}{4\pi}
- \frac{\beta_0\gamma_1-\beta_1\gamma_0}{2\beta_0^2} \frac{\alpha_s(\mu)-\alpha_s(m)}{4\pi}
+ \cdots \right]
\label{sol2}
\end{equation}
The fractional power of $\alpha_s(\mu)/\alpha_s(m)$ in~(\ref{sol2})
contains all leading logs $(\alpha_s L)^n$ in the perturbative series~(\ref{sol1});
the correction inside the brackets contains the subleading logs.
We cannot use the $C_2$ term here until we know $\gamma_2$.

The largest term missing in~(\ref{sol1}) is $\left(\frac{\alpha_s}{4\pi}\right)^3 L^3$.
the largest term missing in~(\ref{sol2}) is $C_2 \left(\frac{\alpha_s}{4\pi}\right)^2$.
Comparing these errors, we can estimate the value of $L$
at which~(\ref{sol2}) becomes a better approximation than~(\ref{sol1}).

The solution~(\ref{sol}) of the renormalization group equation depends on $m$ and $\mu$.
We can rewrite it as a product of a function of $m$ and a function of $\mu$.
Let's subtract and add $\frac{\gamma_0}{2\beta_0}$ inside the integral in~(\ref{sol}).
the difference may be integrated from 0, and we obtain
\begin{align}
&C_m(\mu) = \alpha_s(m)^{\frac{\gamma_0}{2\beta_0}} \hat{C}_m(m) \cdot
\alpha_s(\mu)^{-\frac{\gamma_0}{2\beta_0}} K(\mu)\,,
\label{Nfac}\\
&K(\mu) = \exp \left[ - \int\limits_0^{\alpha_s(\mu)}
\left(\frac{\ga_m(\alpha_s)}{2\beta(\alpha_s)} - \frac{\gamma_0}{2\beta_0}\right)
\frac{d\alpha_s}{\alpha_s} \right]
= 1 - \frac{\beta_0\gamma_1-\beta_1\gamma_0}{2\beta_0^2} \frac{\alpha_s(\mu)}{4\pi} + \cdots
\nonumber\\
&\hat{C}_m(m) = C_m(m) K^{-1}(m) = 1
+ \left(C_1 + \frac{\beta_0\gamma_1-\beta_1\gamma_0}{2\beta_0^2}\right)
\frac{\alpha_s(m)}{4\pi} + \cdots
\nonumber
\end{align}

The most obvious effect of the chromomagnetic interaction is the hyperfine
$B$--$B^*$ splitting:
\begin{equation}
m_{B^*} - m_B = \frac{2}{3 m_b} C_m(\mu) \mu_G^2(\mu)
+ \mathcal{O}\left(\frac{1}{m_b^2}\right)\,,
\label{split1}
\end{equation}
where $\mu_G^2(\mu)$ is the matrix element of the chromomagnetic interaction operator.
The product $C_m(\mu) \mu_G^2(\mu)$ is, of course, $\mu$-independent, and hence
\begin{equation}
\mu_G^2(\mu) = \hat{\mu}_G^2 \alpha_s(\mu)^{\frac{\gamma_0}{2\beta_0}} K^{-1}(\mu)\,.
\label{muGmu}
\end{equation}
The quantity $\hat{\mu}_G^2$ is $\mu$-independent, and hence is equal to
$\Lambda_{\MS}^2$ times some number; we obtain
\begin{equation}
m_{B^*} - m_B = \frac{2}{3 m_b} \alpha_s(m_b)^{\frac{\gamma_0}{2\beta_0}}
\hat{C}_m(m_b) \hat{\mu}_G^2 + \mathcal{O}\left(\frac{1}{m_b^2}\right)\,.
\label{split2}
\end{equation}
We can write a similar equation for $m_{D^*}-m_D$.
The quantities $\hat{\mu}_G^2$ in $b$-quark and $c$-quark HQET differ
by an amount of order $(\alpha_s(m_c)/\pi)^2$ due to decoupling of $c$-quark loops
(Sect.~\ref{SecDec}).
Multiplying~(\ref{split2}) by $m_{B^*}+m_B=2m_b+\mathcal{O}(1/m_b)$
and dividing by a similar $D$-meson equation, we obtain~\cite{ABN}
\begin{align}
\frac{m_{B^*}^2-m_B^2}{m_{D^*}^2-m_D^2}
=& \left(\frac{\alpha_s(m_b)}{\alpha_s(m_c)}\right)^{\frac{\gamma_0}{2\beta_0}}
\left[1 - \left(C_1 + \frac{\beta_0\gamma_1-\beta_1\gamma_0}{2\beta_0^2}\right)
\frac{\alpha_s(m_c)-\alpha_s(m_b)}{4\pi}
+ \mathcal{O}\left(\left(\frac{\alpha_s}{\pi}\right)^2\right) \right]
\nonumber\\
&{} + \mathcal{O}\left(\frac{\Lambda_{\MS}}{m_{c,b}}\right)\,.
\label{splitR}
\end{align}
In the interval between $m_c$ and $m_b$, the relevant number of flavours is $n_f=4$:
\begin{equation*}
\frac{m_{B^*}^2-m_B^2}{m_{D^*}^2-m_D^2}
= \left(\frac{\alpha_s(m_b)}{\alpha_s(m_c)}\right)^{9/25}
\left[ 1 - \frac{7921}{3750} \frac{\alpha_s(m_c)-\alpha_s(m_b)}{\pi} \right]
+ \cdots
\end{equation*}
The experimental value of this ratio is 0.89.
The leading logarithmic approximation gives 0.84;
the NL correction reduces this result by 9\%, giving 0.76.
The agreement is quite good,
taking into account the fact that the $1/m_c$ correction
may be rather large.

\section{Decoupling of heavy quark loops}
\label{SecDec}

Let's consider QCD with $n_l$ light flavours and a single heavy one, say, $c$.
Processes involving only light fields with momenta $p_i\ll m_c$
can be described by an effective field theory --- QCD with $n_l$ flavours.
There are $1/m_c^n$ suppressed local operators in the Lagrangian,
which are the remnants of heavy quark loops shrunk to a point.
This low-energy effective theory is constructed to reproduce
$S$-matrix elements of full QCD expanded to some order in $p_i/m_c$.
Operators of full QCD can be expanded in $1/m_c$
in operators of the effective theory.
Coefficients of these expansions are fixed by matching ---
equating on-shell matrix elements, up to some order in $p_i/m_c$.
Matching full QCD with the low-energy one is called decoupling;
it is very clearly presented in~\cite{CKS},
where references to earlier papers can be found.

In particular, the light fields of full QCD are related
to those of the low-energy theory by
\begin{equation}
q_i = \zeta_q(\mu)^{1/2} q_i'\,,\quad
A_\mu = \zeta_A(\mu)^{1/2} A_\mu'\,,\quad
c = \zeta_c(\mu)^{1/2} c'\,,
\label{DecAq}
\end{equation}
up to $1/m_c$ suppressed terms,
where the fields are renormalized at $\mu$ in both theories.
The coupling constant and gauge-fixing parameter
in both theories are related by
\begin{equation}
g = \zeta_\alpha(\mu)^{1/2} g'\,,\quad
a = \zeta_A(\mu) a'
\label{Decga}
\end{equation}
(we shall see in a moment, why the last coefficient is $\zeta_A$).
It is more convenient to calculate the coefficients $\zeta_i^0$
which relate the bare fields and parameters in both theories.
After that, it is easy to find the renormalized ones:
\begin{equation}
\zeta_q(\mu) = \frac{Z_q'(\mu)}{Z_q(\mu)} \zeta_q^0\,,\quad
\zeta_A(\mu) = \frac{Z_A'(\mu)}{Z_A(\mu)} \zeta_A^0\,,\quad
\zeta_c(\mu) = \frac{Z_c'(\mu)}{Z_c(\mu)} \zeta_c^0\,,\quad
\zeta_\alpha(\mu) = \frac{Z_\alpha'(\mu)}{Z_\alpha(\mu)} \zeta_\alpha^0\,.
\label{DecRen}
\end{equation}

The transverse part of the bare gluon propagator~(\ref{Glue1})
near the mass shell in full QCD is
\begin{equation}
D_{0\bot}(p^2) = \frac{Z_A^{\text{os}}}{p^2}\,,\quad
Z_A^{\text{os}} = \frac{1}{1-\Pi(0)}\,.
\label{DecD}
\end{equation}
In the low-energy theory, $Z_A^{\prime\text{os}}=1$,
because all loop diagrams for $\Pi'(0)$ contain no scale.
Therefore, the matching coefficient $\zeta_A^0$ in the relation $A_0=\zeta_A^0 A_0'$ is
\begin{equation}
\zeta_A^0 = \frac{Z_A^{\text{os}}}{Z_A^{\prime\text{os}}} = \frac{1}{1-\Pi(0)}\,.
\label{DecA0}
\end{equation}
Multiplying by this coefficient also converts $a_0'$ into $a_0$
in the longitudinal part of the propagator.

\begin{figure}[ht]
\begin{center}
\begin{picture}(102,17)
\put(51,8.5){\makebox(0,0){\includegraphics{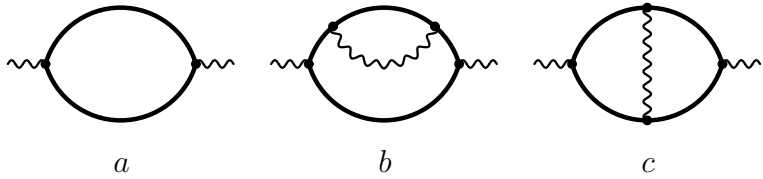}}}
\put(16,-6){\makebox(0,0)[b]{$a$}}
\put(51,-6){\makebox(0,0)[b]{$b$}}
\put(86,-6){\makebox(0,0)[b]{$c$}}
\end{picture}
\end{center}
\caption{Massive loops in gluon self-energy}
\label{DecPi}
\end{figure}

In the full theory, only diagrams with $c$-quark loops (Fig.~\ref{DecPi}) contribute.
It is convenient to extract $\Pi(0)$ using
\begin{equation*}
\Pi(0) = \frac{1}{2d(d-1)} \left.
\frac{\partial}{\partial p^\nu} \frac{\partial}{\partial p_\nu}
\Pi_\mu^\mu(p) \right|_{p=0}\,.
\end{equation*}
In the one-loop approximation (Fig.~\ref{DecPi}a),
$\Pi(0)$ is a combination of one-loop vacuum integrals~(\ref{Tadpole});
we obtain
\begin{equation}
\Pi(0) = - \frac{4}{3} T_F \frac{g_0^2 m_c^{-2\epsilon}}{(4\pi)^{d/2}} \Gamma(\epsilon)
+ \cdots
\label{Pi0}
\end{equation}
Therefore,
\begin{equation*}
\zeta_A^0 = 1 - \frac{4}{3} T_F \frac{g_0^2 m_c^{-2\epsilon}}{(4\pi)^{d/2}} \Gamma(\epsilon)
+ \cdots
\end{equation*}
The ratio of the renormalization constants~(\ref{Z2}) for the gluon field is
\begin{equation*}
\frac{Z_A}{Z_A'} = 1 - \frac{1}{2} \Delta \gamma_{A0} \frac{\alpha_s}{4\pi\epsilon} + \cdots
= 1 - \frac{4}{3} T_F \frac{\alpha_s}{4\pi\epsilon} + \cdots\,,
\end{equation*}
where $\Delta\gamma_{A0}=\frac{8}{3}T_F$ is the contribution of a single flavour
to the one-loop anomalous dimension~(\ref{gammaA}) of the gluon field.
Finally, we obtain the renormalized decoupling constant
\begin{equation}
\zeta_A(\mu) = 1 + \frac{8}{3} T_F \frac{\alpha_s(\mu)}{4\pi} \log\frac{m}{\mu}
+ \cdots\,.
\label{DecA1}
\end{equation}
It is most natural to calculate $\zeta_i(m_c)$ which contain no large logs.
The dependence on $\mu$ can be easily restored using the renormalization group.
It is not too difficult to calculate the two-loop diagrams Fig.~(\ref{DecPi})b, c.
They reduce to combinations of two-loop massive vacuum integrals~(\ref{vac}),
and are proportional to $\Gamma^2(\epsilon)$.
Using also the two-loop anomalous dimension of the gluon field,
we can obtain~\cite{CKS}
\begin{equation}
\zeta_A(m_c) = 1 - \frac{13}{24} T_F (4C_F-C_A) \left(\frac{\alpha_s(m_c)}{4\pi}\right)^2
+ \cdots
\label{DecA2}
\end{equation}

\begin{figure}[ht]
\begin{center}
\begin{picture}(52,20)
\put(26,10){\makebox(0,0){\includegraphics{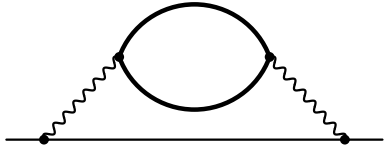}}}
\end{picture}
\end{center}
\caption{Two-loop on-shell massless quark self energy}
\label{OSQCD0}
\end{figure}

Similarly, for the light quark field (see~(\ref{Quark2}))
\begin{equation}
\zeta_q^0 = \frac{Z_q^{\text{os}}}{Z_q^{\prime\text{os}}} = \frac{1}{1-\Sigma_V(0)}\,.
\label{Decq0}
\end{equation}
The only two-loop diagram contributing is Fig.~\ref{OSQCD0}.
Expanding the quark self-energy~(\ref{Sig}) at $m_0=0$ in $p$
up to linear terms, we reduce it to the integrals~(\ref{vac}),
similarly to~(\ref{Sipr})~\cite{CKS,G2}
\begin{equation}
\begin{split}
&\Sigma_V(0) = -i C_F g_0^2 \frac{(d-1)(d-4)}{d}
\int \frac{d^d k}{(2\pi)^d} \frac{\Pi(k^2)}{(k^2)^2}\,,\\
&\zeta_q^0 = \left[ 1 - \Sigma_V(0) \right]^{-1} = 1 +
C_F T_F \frac{g_0^4 m_c^{-4\epsilon}}{(4\pi)^d} \Gamma^2(\epsilon)
\frac{2(d-1)(d-4)(d-6)}{d(d-2)(d-5)(d-7)}\,.
\end{split}
\label{zqos}
\end{equation}
The ratio of the renormalization constants~(\ref{Z2}) for the quark field is
\begin{equation*}
\frac{Z_q}{Z_q'} = 1 + \frac{1}{4}
\left( \gamma_{q0} \Delta \beta_0
+ \frac{1}{2} \Delta \gamma_{A0} \frac{d\gamma_{q0}}{da} a
- \Delta \gamma_{q1} \epsilon \right)
\left(\frac{\alpha_s}{4\pi\epsilon}\right)^2 + \cdots
= 1 + C_F T_F \frac{1}{\epsilon} \left(\frac{\alpha_s}{4\pi}\right)^2 + \cdots\,,
\end{equation*}
where $\Delta\beta_0=-\frac{4}{3}T_F$, $\Delta\gamma_{A0}=\frac{8}{3}T_F$
and $\Delta\gamma_{q1}=-4 C_F T_F$ are the single-flavour contributions.
Finally, we obtain~\cite{CKS}
\begin{equation}
\zeta_q(m_c) = 1 - \frac{5}{6} C_F T_F \left(\frac{\alpha_s(m_c)}{4\pi}\right)^2
+ \cdots
\label{Decq2}
\end{equation}

The decoupling relation for the coupling constant can be derived
from considering any vertex in the theory.
For example, in~\cite{CKS} it was obtained from the ghost-ghost-gluon vertex.
Let $g_0 \Gamma_c$ be the sum of bare one-particle-irreducible ghost-ghost-gluon
vertex diagrams, not including the external propagators.
For this vertex, we have
\begin{equation*}
g_0 \Gamma_c = \left(\zeta_c^0\right)^{-1} \left(\zeta_A^0\right)^{-1/2}
g_0' \Gamma_c'\,,
\end{equation*}
because after multiplication by the three propagators it becomes
the Green function of $c_0$, $\overline{c}_0$, $A_0$.
It is most convenient to nullify all the external momenta,
after a single differentiation in the outgoing ghost momentum.
In the low-energy theory, $\Gamma_c'=1$, if we single out the colour factor.
In full QCD, only diagrams with $c$-quark loops contribute.
They first appear at two loops, and there are three of them.
We have
\begin{equation*}
\zeta_\alpha^0 = \frac{1}{\left(\zeta_c^0\right)^2 \zeta_A^0 \Gamma_c}\,,\quad
\zeta_\alpha = \frac{Z_\alpha'}{Z_\alpha} \zeta_\alpha^0\,.
\end{equation*}
Calculating $\zeta_c$ and $\Gamma_c$, we can obtain the famous result
(see~\cite{CKS})
\begin{equation}
\zeta_\alpha(m_c) = 1 - \left( \frac{4}{9} C_A - \frac{13}{24} C_F \right)
\left(\frac{\alpha_s(m_c)}{4\pi}\right)^2 + \cdots
\label{DecAs}
\end{equation}
It means that at $\mu>m_c$, we have $\alpha_s(\mu)$ whose running is given by
the $\beta(\alpha_s)$ function with $n_l+1$ flavours;
at $\mu<m_c$, we have $\alpha_s'(\mu)$ whose running is given by
$\beta'(\alpha_s')$ with $n_l$ flavours;
at $\mu=m_c$,
\begin{equation*}
\alpha_s(m_c)=\zeta_\alpha(m_c)\alpha_s'(m_c)\,.
\end{equation*}

If we consider $b$-quark HQET instead of QCD, nothing changes.
When all characteristic (residual) momenta become much less than $m_c$,
$c$-quark loops shrink to a point.
From~(\ref{ZHQETOS}), we obtain
\begin{equation}
\widetilde{\zeta}_Q^0 = \frac{\Z_Q^{\text{os}}}{\Z_Q^{\prime\text{os}}}
= 1 - C_F T_F \frac{g_0^4 m_c^{-4\epsilon}}{(4\pi)^d} \Gamma^2(\epsilon)
\frac{2(d-1)(d-6)}{(d-2)(d-5)(d-7)}\,.
\label{DecQ0}
\end{equation}
The ratio of renormalization constants is, from~(\ref{gammaQ2}),
\begin{equation*}
\frac{\Z_Q}{\Z_Q'} = 1 + 2 C_F T_F \left(\frac{\alpha_s}{4\pi\epsilon}\right)^2
\left(1 - \frac{4}{3}\epsilon\right)\,.
\end{equation*}
Finally, we obtain~\cite{G2}
\begin{equation}
\widetilde{\zeta}_Q(m_c) = 1 + \frac{52}{9} C_F T_F \left(\frac{\alpha_s(m_c)}{4\pi}\right)^2
+ \cdots
\label{DecQ2}
\end{equation}

Finally, we shall discuss decoupling of $c$-quark loops in the $b$-quark
chromomagnetic interaction constant.
Both HQET with $c$-loops and the effective low-energy theory
must give identical on-shell scattering amplitudes,
up to corrections suppressed by powers of $1/m_c$.
Therefore, from~(\ref{MatchM}) we obtain
\begin{equation}
C_m(\mu) = \frac{\Z_m(\mu)}{\Z_m'(\mu)} \frac{\widetilde{\mu}'}{\widetilde{\mu}} C_m'(\mu)\,.
\label{DecCm}
\end{equation}
Here $\widetilde{\mu}'=1$, and $\widetilde{\mu}$ is given by~(\ref{HQETmug}),
with just the $c$-quark contribution.
The ratio of the renormalization constants $\Z_m$~(\ref{Z2}) is
\begin{equation}
\frac{\Z_m}{\Z_m'} = 1 + \frac{1}{4}
\left( \gamma_0 \Delta \beta_0 - \Delta \gamma_1 \epsilon \right)
\left(\frac{\alpha_s}{4\pi\epsilon}\right)^2 + \cdots
= 1 - C_A T_F \left( \frac{2}{3} - \frac{13}{8} \epsilon \right)
\left(\frac{\alpha_s}{4\pi\epsilon}\right)^2 + \cdots\,,
\label{DecZ}
\end{equation}
where $\Delta\beta_0=-\frac{4}{3}T_F$ and $\Delta\gamma_1=-\frac{4}{9}\cdot 13 C_A T_F$
are the single-flavour contributions.
Re-expressing~(\ref{HQETmug}) via $\alpha_s(m_c)$ and expanding in $\epsilon$,
we see that singular terms cancel, and
\begin{equation}
C_m(m_c) = \left[ 1 + \frac{71}{27} C_A T_F \left(\frac{\alpha_s(m_c)}{4\pi}\right)^2
+ \cdots \right] C_m'(m_c)\,.
\label{DecCmRes}
\end{equation}
Therefore, when crossing $\mu=m_c$, we have to adjust $C_m(\mu)$
according to~(\ref{DecCmRes}),
and at $\mu<m_c$ the renormalization-group running is driven
by the $\beta$ function and the anomalous dimension with $n_f=3$.

\section{Conclusion}
\label{SecConc}

Part 2 of these lectures will deal with heavy-light quark currents,
and with renormalons in HQET.
More physics applications of HQET can be found in other reviews.
Here is a partial list, in no particular order: \cite{MW}--\cite{Srev}.

I am grateful to K.~G.~Chetyrkin, who proposed the idea to give these lectures;
to Th.~Mannel, J.~H.~K\"uhn, T.~van~Ritbergen for useful discussions
during my stay in Karlsruhe;
to D.~J.~Broadhurst, M.~Neubert, A.~Czarnecki, A.~I.~Davydychev, O.~I.~Yakovlev
for collaboration on various HQET projects;
to D.~V.~Shirkov, D.~I.~Kazakov, S.~V.~Mikhailov, A.~P.~Bakulev
for organizing the very interesting Calc-2000 school/workshop in Dubna.


\end{document}